\RequirePackage{fix-cm}
\documentclass[twocolumn]{svjour3}          
\smartqed  
\usepackage{graphicx}
	\usepackage{amsmath,latexsym, graphics,amsfonts,fullpage,bbm}
	\usepackage[bookmarks,colorlinks=true,linkcolor=black]{hyperref}
	\usepackage{algpseudocode,algorithm,color}
	\usepackage{natbib}
\usepackage{subcaption}
\def\fixedlabel#1#2{%
  \@bsphack
  \begingroup
    \@onelevel@sanitize\@currentlabelname
    \edef\@currentlabelname{%
      \expandafter\strip@period\@currentlabelname\relax.\relax\@@@%
    }%
    \phantomsection%
    \protected@write\@auxout{}{%
      \string\newlabel{#1}{%
        {#2}%
        {\thepage}%
        {#2}%
        {\@currentHref}{}%
      }%
    }%
  \endgroup
  \@esphack
}

\begin{document}

\title{Sequentially Constrained Monte Carlo
}


\author{Shirin Golchi         \and
        David A. Campbell 
}


\institute{S. Golchi \at
              Applied Statistics Centre, Columbia University, New York, NY 10027\\
              \email{sg3252@columbia.edu}           
           \and
           D. Campbell \at
           Department of Statistics and Actuarial Science, Simon Fraser University, 13450 102nd Avenue,
Surrey, BC, V3T 0A3
}

\date{}

\maketitle

\begin{abstract}
Constraints can be interpreted in a broad sense as any kind of explicit restriction over the parameters.  While some constraints are defined directly on the parameter space, when they are instead defined by known behavior on the model, transformation of constraints into features on the parameter space may not be possible. Difficulties in sampling from the posterior distribution as a result of incorporation of constraints into the model is a common challenge leading to truncations in the parameter space and inefficient sampling algorithms. We propose a variant of sequential Monte Carlo algorithm for posterior sampling in presence of constraints by defining a sequence of densities through the imposition of the constraint.  Particles generated from an unconstrained or mildly constrained distribution are filtered and moved through sampling and resampling steps to obtain a sample from the fully constrained target distribution.  General and model specific forms of constraints enforcing strategies are defined.  The Sequentially Constrained Monte Carlo algorithm is demonstrated on constraints defined by monotonicity of a function, densities constrained to low dimensional manifolds, adherence to a theoretically derived model, and model feature matching.
\keywords{Particle Filtering \and Constraints \and Differential Equation Models \and Parameter Estimation \and  Approximate Bayesian Computation \and Sequential Monte Carlo }
\end{abstract}

\section{Introduction}
\label{intro}
Constraints are tools to incorporate external information and ensure parameters remain interpretable.  In non-linear or high dimensional models, parameter constraints are complex to define, even when constraints on the model expectation surface may remain simple.  Constraints can be defined in a broad sense as any explicit restriction over the parameter or model space.  A few examples are: inequality constraints over model parameters to enforce physical laws such as conservation of mass and energy; ensuring monotonicity or convexity of functions in regression or non-parametric smoothing; conforming to theoretical behavior governed by a model.

Imposition of constraints can induce multi-modality and/or zero probability regions resulting in challenges in sampling random variable $\boldsymbol{\theta}$ from a target distribution and in the Bayesian context may also lead to disagreements between the prior and likelihood.   In this paper we extend the utility of Sequential Monte Carlo (SMC) samplers \cite{Moral06} by defining a sequence of distributions by their enforcement of a constraint through the proposed Sequentially Constrained Monte Carlo\\ (SCMC) algorithm.  We  connect a ``simple" distribution, $\pi_0(\boldsymbol{\theta})$ to the target distribution, $\pi_T(\boldsymbol{\theta})$,  via a path defined by the strictness of constraint enforcement, thereby generalizing the usual transitions of SMC between $\pi_0(\boldsymbol{\theta})$ and $\pi_T(\boldsymbol{\theta})$.  Furthermore, we show general applicability of SCMC by creatively defining constraints. 

To showcase SCMC, we begin with the toy problem of polynomial regression on noisy observations of monotone functions. Sequentially imposing the monotonicity information by defining a ``soft'' positivity constraint over the derivative polynomial produces more accurate predictions while satisfying the monotonicity constraints. The second example involves sampling a bivariate density constrained to take non-zero probabilities only when both variables lie on lower dimensional manifolds.  In our third application we estimate a mixture of discrete and continuous parameters of an ordinary differential equation model where we generalize the usual definition of constraint to include model adherence. In this case, full constraint enforcement produces multiple disjoint modes.  The fourth example focuses on parameter estimation for a chaotic stochastic differential equation model where we define the constraint through a form of sequentially expanding model adherence criteria in an ABC algorithm.  The first two applications define a general strategy for enforcing constraints, whereas the last two constraints showcase problem specific strategies.
		
The rest of the paper is organized as follows; in Section \ref{sec:SMC-ref}, we provide a background on SMC samplers and the commonly used versions of it. We explain the  choice of the sequence of densities that outlines the SCMC in Section \ref{scmc-par}. In Section \ref{sec:probit_general}, a general strategy is defined through a probit based soft constraint model.  This SCMC strategy is applied to the monotone polynomial regression model and a 1 dimensional model embedded on a nonlinear manifold in the 2 dimensional sampling space while avoiding the need for a Jacobian of the transformation.  Some application specific constraint strategies are introduced in Section \ref{sec:app_constraints}.  Here Parameter estimation for ODE models in a sequential framework is explained in Section \ref{sec:ode}. In Section \ref{sec:abc}, the Sequentially Constrained Approximate Bayesian Computation (SCMC ABC) algorithm is introduced and Section \ref{sec:conclusion} follows with concluding remarks.

\section{Sequential Monte Carlo}\label{sec:SMC-ref}

	SMC samplers are a family of algorithms that can be used in many challenging scenarios where random walk Markov chain Monte Carlo (MCMC) methods fail in efficiently sampling $\boldsymbol{\theta}$ from its target distribution. SMC algorithms take advantage of a sequence of bridging distributions that bridge between $\pi_0(\boldsymbol{\theta})$, a distribution that is straightforward to sample, and $\pi_T(\boldsymbol{\theta})$, a difficult to sample target distribution.  In Bayesian inference, these distributions are typically the prior $\pi_0(\boldsymbol{\theta})$ and posterior $\pi_T(\boldsymbol{\theta})$. 
	
     SMC discretizes a sequence of densities   	\begin{equation*}
	\label{pi-tau}
\left\{	\pi_{t}(\boldsymbol{\theta})=\frac{\eta_{t}(\boldsymbol{\theta})}{Z_{t}}\right\}_{t=0}^{T}
	\end{equation*} 
\noindent	between $\pi_0(\boldsymbol{\theta})$ and $\pi_T(\boldsymbol{\theta})$ with possibly unknown normalizing constant  $Z_{t}$ and kernel $\eta_{t}$ which can be evaluated for a given $\boldsymbol{\theta}$.
The initial sample of particles, $\boldsymbol{\theta}_0^{1:N}\sim\pi_0(\boldsymbol{\theta})$ are filtered through iterative jittering and importance resampling steps to eventually obtain a sample from $\pi_T(\boldsymbol{\theta})$ as outlined in Algorithm~\ref{SMC-original}  \cite{Moral06}.

	
Algorithm~\ref{SMC-original}, is very general in the sense that many possible choices could be made for the inputs of the algorithm. The choice of the inputs, especially the forward kernels used for jittering the sample, $K_t(\cdot)$,  and the backward kernels, $L_t$, that ensure the weights are defined according to the posterior at time $t$, can change the order of the steps in Algorithm~\ref{SMC-original}. A variety of options for the forward and backward kernels and the resulting expressions for the incremental weights, $w_i$, are provided by \cite{Moral06}. In the following, we explain the specific choices that are made for all our examples.
	
	\begin{algorithm}[t]
	\caption{Sequential Monte Carlo Sampler}\label{SMC-original}
	\begin{algorithmic}[1]
	\renewcommand{\algorithmicrequire}{\textbf{Input:}}
	\renewcommand{\algorithmicensure}{\textbf{Return:}}
	\Require Forward and backward kernels, $K_t$ and $L_t$.\\
		
	Generate an initial sample $\boldsymbol{\theta}^{1:N}_0\sim \pi_0$;\\
	$W_0^j\gets \frac{1}{N}$, $j=1,\ldots,N$;
	\For{$t:=1, \ldots, T$} 
	\begin{itemize}
		
	\If{$\text{ESS}=\left(\sum_{j=1}^N\left(W_{t-1}^j\right)^2\right)^{-1}<\frac{N}{2}$} \State{resample $\boldsymbol{\theta}^{1:N}_{t-1}$ with weights $W_{t-1}^{1:N}$} \State{$W_{t-1}^{1:N}\gets \frac{1}{N}$}
	
	 \EndIf
	\item Sample $\boldsymbol{\theta}_{t}^{1:N}\sim K_{t}$;
	\item $W_t^j\gets W_{t-1}^j w_t^j$ where $w_t^j=\frac{\eta_{t}(\boldsymbol{\theta}_{t}^j)L_{t-1}(\boldsymbol{\theta}_{t}^j,\boldsymbol{\theta}_{t-1}^j)}{\eta_{t-1}(\boldsymbol{\theta}_{t-1}^j)K_{t}(\boldsymbol{\theta}_{t-1}^j,\boldsymbol{\theta}_{t}^j)}$, $j=1,\ldots,N$;
		
	\item Normalize $W_{t}^{1:N}$.
		
	\end{itemize}
	\EndFor
		
	\Ensure Particles $\boldsymbol{\theta}_{1:T}^{1:N}$.
	\end{algorithmic}
	\end{algorithm}

	At algorithmic stage $t$, the forward kernel, $K_{t}$  is chosen to be a MCMC kernel of invariant distribution $\pi_{t}$. The associated backward kernel recommended by \cite{Moral06} for this choice of $K_{t}$ is
	\begin{equation*}
	L_{t-1}(\theta_t,\theta_{t-1})=\frac{\pi_{t}(\boldsymbol{\theta}_{t-1})K_{t}(\boldsymbol{\theta}_{t-1},\boldsymbol{\theta}_{t})}{\pi_{t}(\boldsymbol{\theta}_{t})}.
	\end{equation*} 
	The above backward kernels are referred to as the ``sub-optimal" by \cite{Moral06} since they are obtained by replacing the marginal importance distributions that do not have a closed form representation by $\pi_t$ in the optimal backward kernels. This choice of the forward and backward kernels results in the simplified form of the incremental weights,
	\begin{equation*}
	w_{t}^j=\frac{\eta_{t}(\boldsymbol{\theta}^j_{t-1})}{\eta_{t-1}({\boldsymbol{\theta}^j_{t-1}})},
	\end{equation*}
	which means that the resampling weights, $W_{t}^{1:N}$, are independent of $K_t$ and $L_t-1$. By postponing sampling from $K_t$ in Algorithm~\ref{SMC-original} until after the $w_{t}^j$ are evaluated and particles are resampled we obtain  Algorithm~\ref{SMC-thesis}.   Algorithm~\ref{SMC-thesis} is the generic algorithm that is used as a basis for all the algorithms tailored to our examples.

	\begin{algorithm}[t]
	\caption{Sequential Monte Carlo}\label{SMC-thesis}
	\begin{algorithmic}[1]
	\renewcommand{\algorithmicrequire}{\textbf{Input:}}
	\renewcommand{\algorithmicensure}{\textbf{Return:}}
	\Require MCMC transition kernels $K_{t}$.\\
	
	Generate an initial sample $\boldsymbol{\theta}^{1:N}_0\sim \pi_0$;\\
	$W_1^{1:N}\gets \frac{1}{N}$;
	\For{$t:=1, \ldots, T$} 
	\begin{itemize}

	 \item $W_t^j\gets W_{t-1}^j w_t^i$ where $w_t^j=\frac{\eta_{t}(\boldsymbol{\theta}_{t-1}^j)}{\eta_{t-1}(\boldsymbol{\theta}_{t-1}^j)}$, $j=1,\ldots,N$;
           \item Normalize $W_{t}^{1:N}$.
	
	           \If{$\text{ESS}=\left(\sum_{j=1}^N\left(W_{t-1}^j\right)^2\right)^{-1}<\frac{N}{2}$} \State{resample $\boldsymbol{\theta}^{1:N}_{t-1}$ with weights $W_{t-1}^{1:N}$;} \State{$W_{t-1}^{1:N}\gets \frac{1}{N}$;}
	
	 \EndIf
	
	\item Sample $\boldsymbol{\theta}_{t}^{1:N}\sim K_{t}$;
	
	\end{itemize}
	\EndFor
	
	\Ensure Particles $\boldsymbol{\theta}_{1:T}^{1:N}$.
	\end{algorithmic}
	\end{algorithm}


The main advantage of SMC over MCMC is the facility of embarrassingly parallel computation in the time consuming steps of the algorithm.  For example weight calculation and Metropolis Hastings jittering via the forward kernel is performed independently on individual particles allowing the particles to be split into batches for parallel computation.

	A potential problem that can break SMC is particle degeneracy.  This term describes a state in which all but a few particles acquire small or zero resampling weights, $W^j_t$, and therefore few unique particles $\boldsymbol{\theta}^j_t$ are resampled and filtered into the next step of the sequence. The discrepancy between two consecutive distributions plays a role in particle degeneracy.  When the difference between $\pi_t$ and $\pi_{t+1}$, is small, there is little chance of filtering a large number of particles out via negligible values of $W^j_t$. As an additional fix to particle degeneracy, particles are jittered by the forward kernel, $K_t$, moving the particles under the current distribution.  However, as sampling from $K_t$ is typically performed by Metropolis-Hastings, an adequate acceptance rate is necessary to prevent particle degeneracy.  


	\section{Sequences of Densities}
	\label{scmc-par}
	
The key component of SMC is the sequence of distributions through which the particles evolve towards the target distribution. The length of the sequence must be tuned to the dimensionality of the problem for SMC to be stable \citep{Beskos14}. The two main SMC sequences of bridging distributions are based on gradually introducing the likelihood in the posterior. Starting from a sample generated from $\pi_0(\boldsymbol{\theta})$, for the vector of parameters, $\boldsymbol{\theta}$, parameter values are filtered into samples from the posterior distribution, $\pi_T(\boldsymbol{\theta})=\pi(\boldsymbol{\theta}\mid \mathbf{y})$, with data, $\mathbf{y}$. In the first approach, the posterior is the target distribution of interest and the likelihood is tempered with a sequence of temperature parameters, $0=\tau_0<\tau_1<\ldots<\tau_T=1$, giving rise to a power posterior,
	\begin{equation}
	\pi_t(\boldsymbol{\theta})=\pi_t(\boldsymbol{\theta} \mid \mathbf{y})\propto P(\mathbf{y}\mid \boldsymbol{\theta})^{\tau_t} \pi_0(\boldsymbol{\theta}) \label{powerposterior}
	\end{equation}
	and associated sequence $\{\pi_t(\boldsymbol{\theta}\mid \mathbf{y})\}_{t=0}^{T}$ for the SMC algorithm \citep{Moral06}.	The second likelihood induction method, often referred to as particle filtering, has a natural discretization where, in this case, the parameter defining the sequence, $\tau$, is used to denote inclusion of the first $\tau$ elements of the data vector $\mathbf{y}$. The  $t^{\mbox{th}}$ sequential distribution, where $0= \tau_0 \leq \tau_1\leq\ldots \leq \tau_T =N$ is given by:  
	\begin{equation}\begin{array}{rl}
	\pi_t(\boldsymbol{\theta} \mid \mathbf{y}) 	\propto& P(\mathbf{y}_1,\ldots,\mathbf{y}_{\tau_t}\mid \boldsymbol{\theta}) \pi(\boldsymbol{\theta}) \\
	=& P(\mathbf{y}_{\tau_t}\mid \boldsymbol{\theta}) P(\mathbf{y}_1,\ldots,\mathbf{y}_{\tau_t-1}\mid\boldsymbol{\theta}) \\
	\propto& P(\mathbf{y}_{\tau_t}\mid \boldsymbol{\theta}) \pi_{\tau_t-1}(\boldsymbol{\theta} \mid \mathbf{y}).
	\end{array}\label{eq:particlefilter}
	\end{equation}	
This case works well for online estimation where data is available sequentially.  The posterior defined by the inclusion of all of the current data becomes the prior for the next stage of the algorithm where more data becomes available.  At each stage particles are moved towards the target posterior while the target itself shifts at the next stage due to the inclusion of new data \citep{Chopin02,DoucetBook}.

In order for us to be able to define suitable bridging distributions, the features of the target distribution that create challenges in sampling need to be investigated. We consider the case that imposition of a constraint on the model is the factor responsible for difficulties faced in sampling from the target distribution. The novelty of our approach is in the way that the sequence, $\{\pi_t\}_{t=0}^{T}$, is defined. The sequence of densities is constructed by relaxation of a constraint, either fully or partially, such that sampling is feasible. Suppose that $\tau$ is a tuning parameter that controls the rigidity of the constraint incorporated into the model. We define the $t^{\text{th}}$ distribution in the sequence as
	\begin{equation*}
	\pi_{t}(\theta)=\pi(\theta;\tau=\tau_{t}).
	\end{equation*}
	Suppose that by increasing $\tau$ the constraint is more strictly imposed and $\tau=\tau_T$ assures full imposition of the constraint. The sequence of densities is therefore determined by an increasing schedule over the constraint parameter, $\tau$,
	\begin{equation*}
	\tau_0<\tau_1<\ldots<\tau_T.
	\end{equation*}
The proposed SCMC algorithm can be used as long as the strictness of the model constraints can be systematically increased to construct the sequence of densities.  Defined with this sequence, the associated incremental weights $w_{t}^j$ and backwards kernel $L_{t-1}(\theta_t,\theta_{t-1})$ used in Algorithms~\ref{SMC-original} and \ref{SMC-thesis} ensure detailed balance is met and algorithmic convergence is subject to the usual SMC considerations \citep{Moral06}. 

We note that the approach taken by \cite{Moffa14} fits into this framework: they sample from truncated multivariate normal distributions using a sequence of multivariate non-central $t$-distributions with increasing degrees of freedom. We showcase a general constraint enforcement strategy as well as application specific strategies.   general form of constraint where the posterior samples can be made arbitrarily close to the fully constrained target but in practice the fully constrained limit may be unattainable is outlined in Section \ref{sec:probit_general} with 2 examples.  Two application specific constraints where the constraint can be fully enforced are given in Section \ref{sec:app_constraints}.

\section{Probabilistic Constraints}\label{sec:probit_general}
Soft constraints are relaxations where the limiting case as $\tau\rightarrow\infty$ ensures complete compliance, however finite resources may leave the samples arbitrarily close but not completely attaining the constraint.  Probabilistic constraints ensure that the probability of exceeding a constraint by some controlled tolerance is low.
Consider the target distribution for random variable $\theta$ subject to constraining  $\theta\in \mathcal{A}$, described through the indicator function $I_\mathcal{A}(\theta)$:
\begin{equation}\pi(\theta) \propto f(\theta)I_\mathcal{A}(\theta).\label{eq:general}\end{equation}
A general strategy for defining constraints is replacing $I(\theta)$ in (\ref{eq:general}) with a probabilistic function that takes values close to 1 for samples where $\theta\in \mathcal{A}$ and takes values close to 0 depending on how far a point $\theta$ is from being in $\mathcal{A}$. The slope of this probabilistic function is controlled by $\tau$. As a result, $\tau$ increases to enforce the constraint by mimicking the indicator function at large values of $\tau$ but leaving the samples unconstrained at low value of $\tau$. The probabilistic constraint acts as a distance metric $||\theta||^{\tau}_\mathcal{A}$ between a point $\theta$ and the nearest boundary of set $\mathcal{A}$, where $\lim_{\tau \rightarrow \infty}||\theta||_\mathcal{A}^{\tau}=I_\mathcal{A}(\theta)$ such that the $t^{\text{th}}$ stage target distribution in the  SCMC algorithm is
	\begin{align}\label{eq:probit-constraint}
	\pi_t\left(\theta \right) \propto f(\theta) ||\theta||_\mathcal{A}^{\tau},
	\end{align}
leaving (\ref{eq:general}) as the limiting distribution.  We demonstrate the probabilistic constraint through two examples, the first has a constraint on the expectation surface and the second has a constraint defined directly on the random variable.


\subsection{Monotone Polynomial Regression}
\label{scmc-toy}
	In this section we use the SCMC algorithm to model noisy observations from a monotone function. We fit a fixed order polynomial regression model to data generated from monotone increasing functions. While a rich literature exists on monotone functional inference  \cite[see for example][]{Ramsay98, He98, Dette06}, the purpose of this section is to exemplify the adaptation of the SCMC algorithm in a simple framework to help understand the implementation and the effectiveness of sequentially constraining the model.
	
Let the data, $\mathbf{y}=\left(y_1,\ldots,y_n\right)^T$, be noisy observations of a monotone function, $f$, at\\ $\mathbf{x}=\left(x_1,\ldots,x_n\right)^T$. With no loss of generality, suppose that $x_i\in [0,1]$. Consider a $p^{\text{th}}$-order polynomial regression model,
	\begin{equation*}
	\mathbf{y}=\mathbf{X}\boldsymbol{\beta}+\boldsymbol{\epsilon},
	\end{equation*}
	where 
	\begin{equation*}
	\mathbf{X}=\begin{pmatrix}
	  \mathbf{1} & \mathbf{x} & \mathbf{x}^2 &\cdots & \mathbf{x}^p \\
	 \end{pmatrix};
	\end{equation*}
	and $\boldsymbol{\epsilon}=\left(\epsilon_1,\ldots,\epsilon_n\right)^T$ is the vector of independent and identically distributed mean-zero normal random errors with variance $\sigma^2$. 

We make inference about the coefficients, $\boldsymbol{\beta}$, and the variance parameter, $\sigma^2$, while constraining the first derivative, $\frac{\partial}{\partial{x}}\mathbf{X}\boldsymbol{\beta}$, to be positive, for $x\in[0,1]$. Using the prior distribution, $\pi\left(\boldsymbol{\beta},\sigma^2\right)$, the target posterior distribution given the data and the monotonicity constraint is given by,
	\begin{equation}
	\label{target-post}
	\pi\left(\boldsymbol{\beta},\sigma^2\mid \mathbf{X},\mathbf{y},\frac{\partial\mathbf{X}\boldsymbol{\beta}}{\partial{x}}>0\right).
	\end{equation}
	Here the set $\mathcal{A}:=\left\{(\boldsymbol{\beta},\sigma^2): \frac{\partial\mathbf{X}\boldsymbol{\beta}}{\partial{x}}>0\right\}$ in (\ref{eq:probit-constraint}).  To sample from (\ref{target-post}), we define a sequence of distributions for the parametrization of the constraint that admits (\ref{target-post}) as its limit. Using a probit function that adds the monotonicity information to the posterior distribution \citep{RV2010} we define:
	\begin{equation*}
	\label{distance-metric-monotone}
	||(\boldsymbol{\beta},\sigma^2)||_\mathcal{A}^{\tau}:=\prod_{i=1}^n \Phi\left(\tau\frac{\partial\mathbf{X}\boldsymbol{\beta}}{\partial{x}}|_{x=x_i}\right).
	\end{equation*}
where $\Phi\left(\cdot\right)$ is the standard normal cumulative distribution function. Therefore,
	\begin{align}\label{eq:probit}
	\pi\left(\boldsymbol{\beta},\sigma^2\mid \mathbf{X},\mathbf{y},\tau\right)\propto &\pi(\boldsymbol{\beta},\sigma^2){\cal N}(\mathbf{y}-\mathbf{X}\boldsymbol{\beta}; \mathbf{0},\sigma^2\mathbf{I})\nonumber\\ &\prod_{i=1}^n \Phi\left(\tau\frac{\partial\mathbf{X}\boldsymbol{\beta}}{\partial{x}}|_{x=x_i}\right),
	\end{align}
As $\tau \rightarrow \infty$, the term $ \Phi\left(\tau\frac{\partial\mathbf{X}\boldsymbol{\beta}}{\partial{x}}|_{x=x_i}\right)$ becomes an indicator function taking a value of 1 only if the derivative is positive at the observation point $x_i$.   Positive values of the derivatives at a finite set of points does not guarantee monotonicity in general; however, since polynomials are smooth functions restricting the derivatives at the values of $\mathbf{x}$ to be positive will normally impose monotonicity as long as $\mathbf{x}$ is not too sparse. Consequently (\ref{eq:probit}) converges to (\ref{target-post}) as $\tau \rightarrow \infty$.

The above parametrization of the monotonicity constraint defines the sequence of distributions, $\{\pi_{t}\}_{t=0}^{T}$,
	\begin{equation*}
	\pi_t\equiv \pi\left(\boldsymbol{\beta},\sigma^2\mid \mathbf{X},\mathbf{y},\tau_t\right),
	\end{equation*}
	with an increasing sequence of monotonicity parameters, 
	\begin{equation*}
	0=\tau_0<\tau_1<\ldots<\tau_T\rightarrow \infty.
	\end{equation*}
	The incremental weights in the SCMC implementation of  Algorithm \ref{SMC-thesis} simplifies to 
	\begin{equation*}
	w_{t}^j=\frac{\prod_{i=1}^n \Phi\left(\tau_{t}\frac{\partial}{\partial{x}}\mathbf{X}\boldsymbol{\beta}^{j}_{t-1}|_{x=x_i}\right)}{\prod_{i=1}^n \Phi\left(\tau_{t-1}\frac{\partial}{\partial{x}}\mathbf{X}\boldsymbol{\beta}^{j}_{t-1}|_{x=x_i}\right)},
	\end{equation*}
	where the likelihood does not need to be evaluated in order to calculate the incremental weights. 

With the choice of conjugate prior distributions, the posterior distribution in the unconstrained case ($\tau=0$) can be obtained analytically \citep{Box73}  facilitating sampling from $\pi_0$ at the first step of the algorithm. 
	
\begin{figure*}[ht]
  \centering
  \begin{tabular}{c@{\quad}ccc}
    & $\tau=0$ & $\tau=1$ & $\tau=10^5$\\
    \rotatebox{90}{\hskip 15pt  polynomial ($f_1$)} & \includegraphics[width=.3\textwidth]{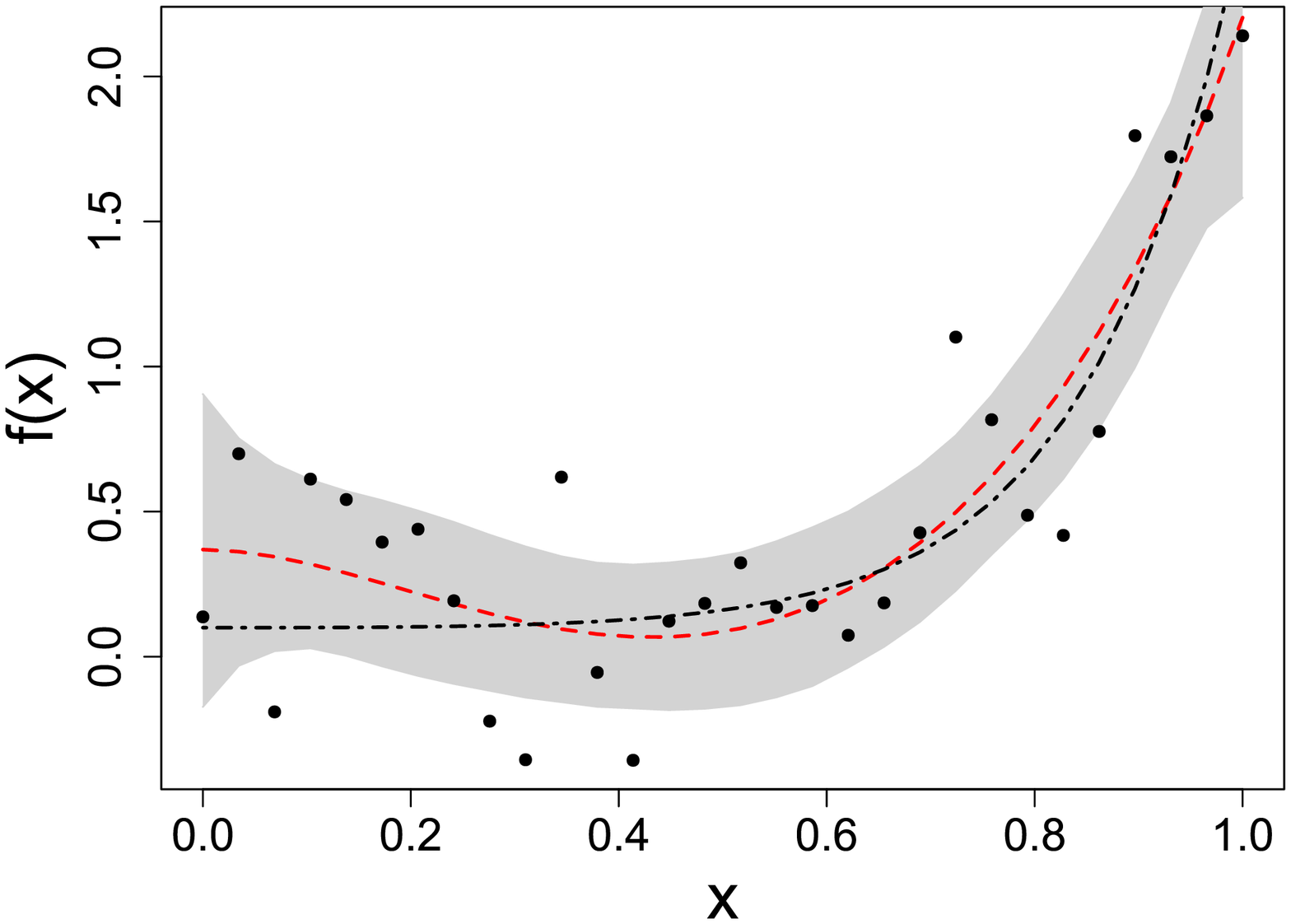} 
      & \includegraphics[width=.3\textwidth]{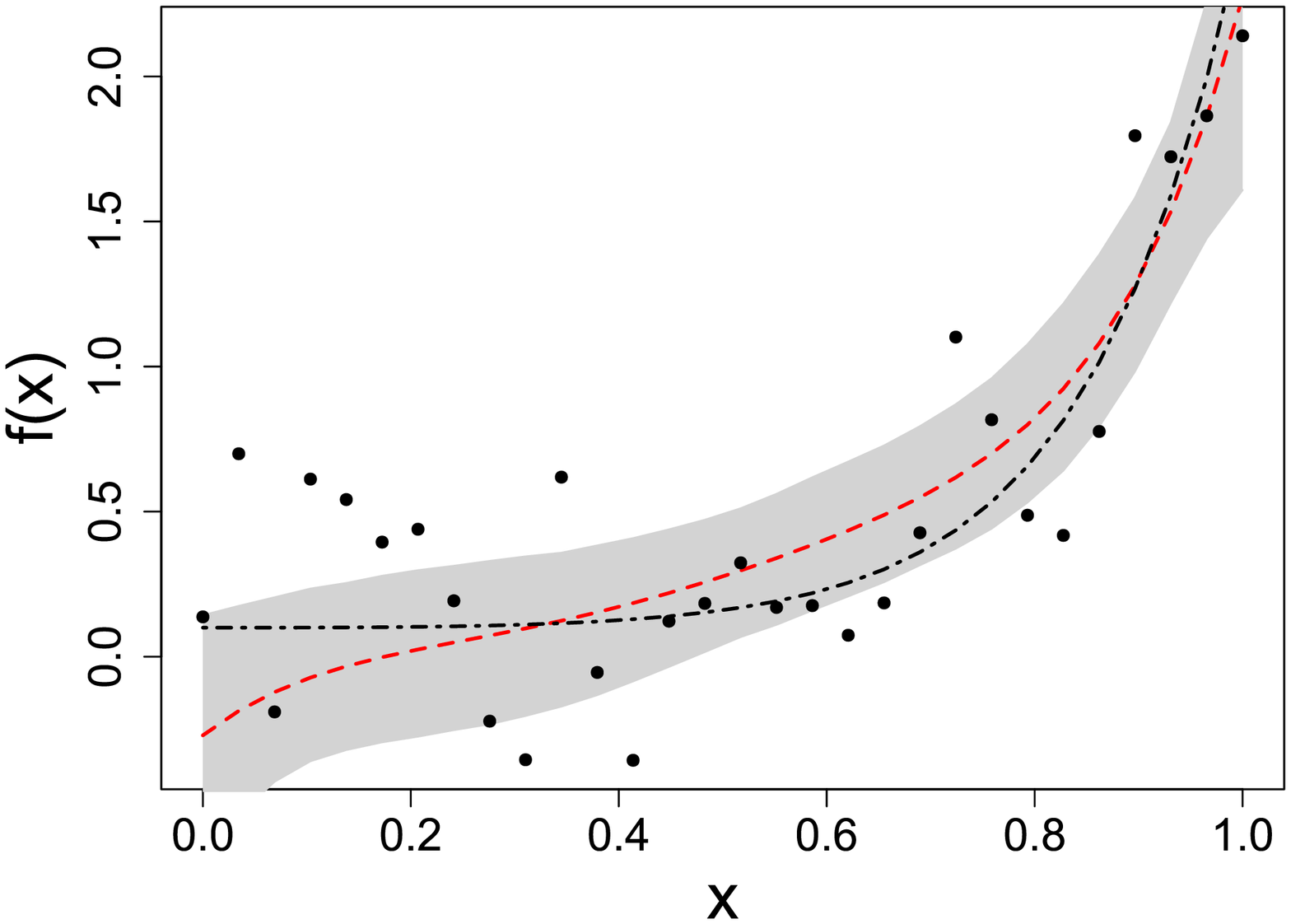}
 &\includegraphics[width=.3\textwidth]{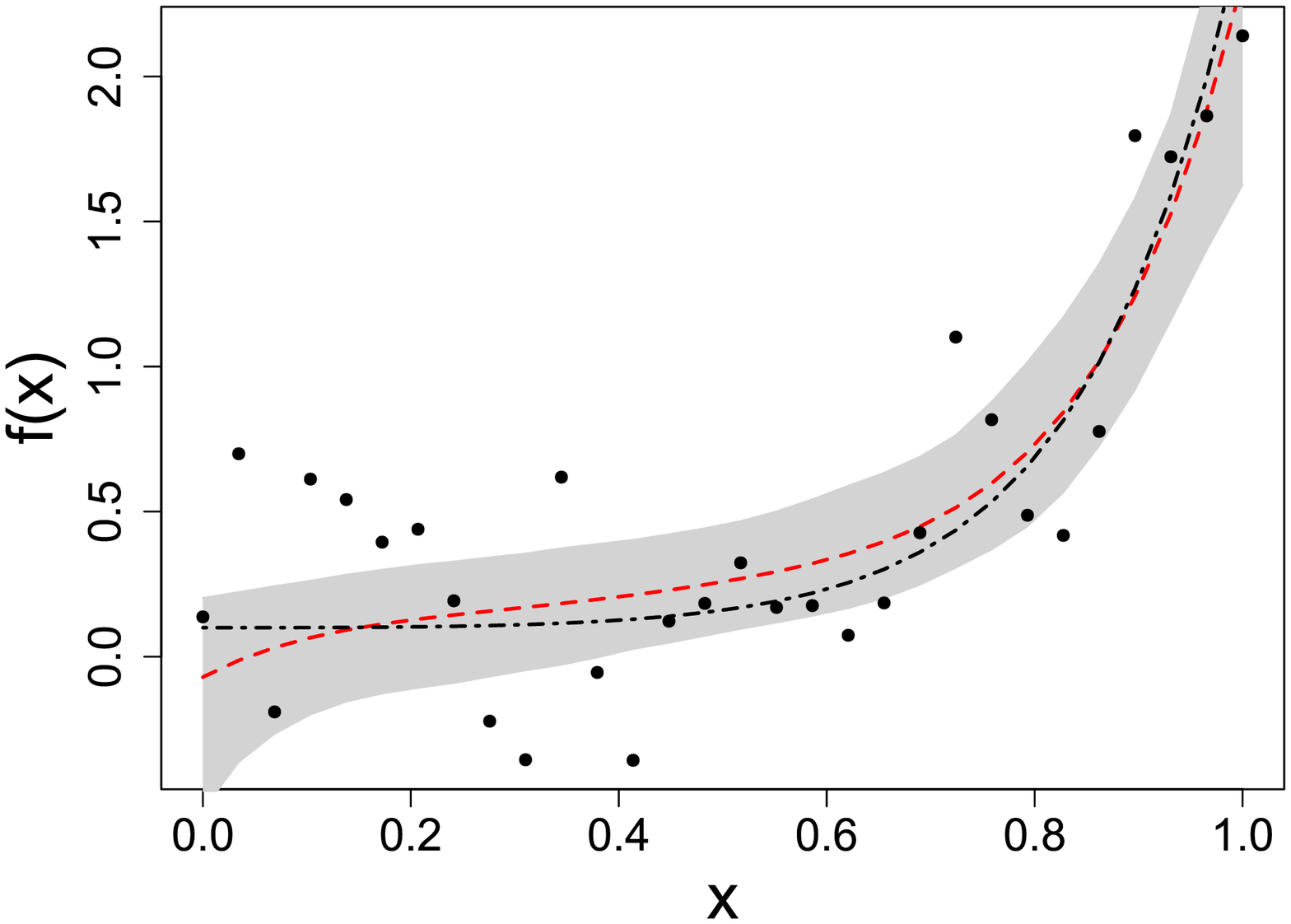}\\ 
    \rotatebox{90}{\hskip 15pt  logarithmic ($f_2$)} & \includegraphics[width=.3\textwidth]{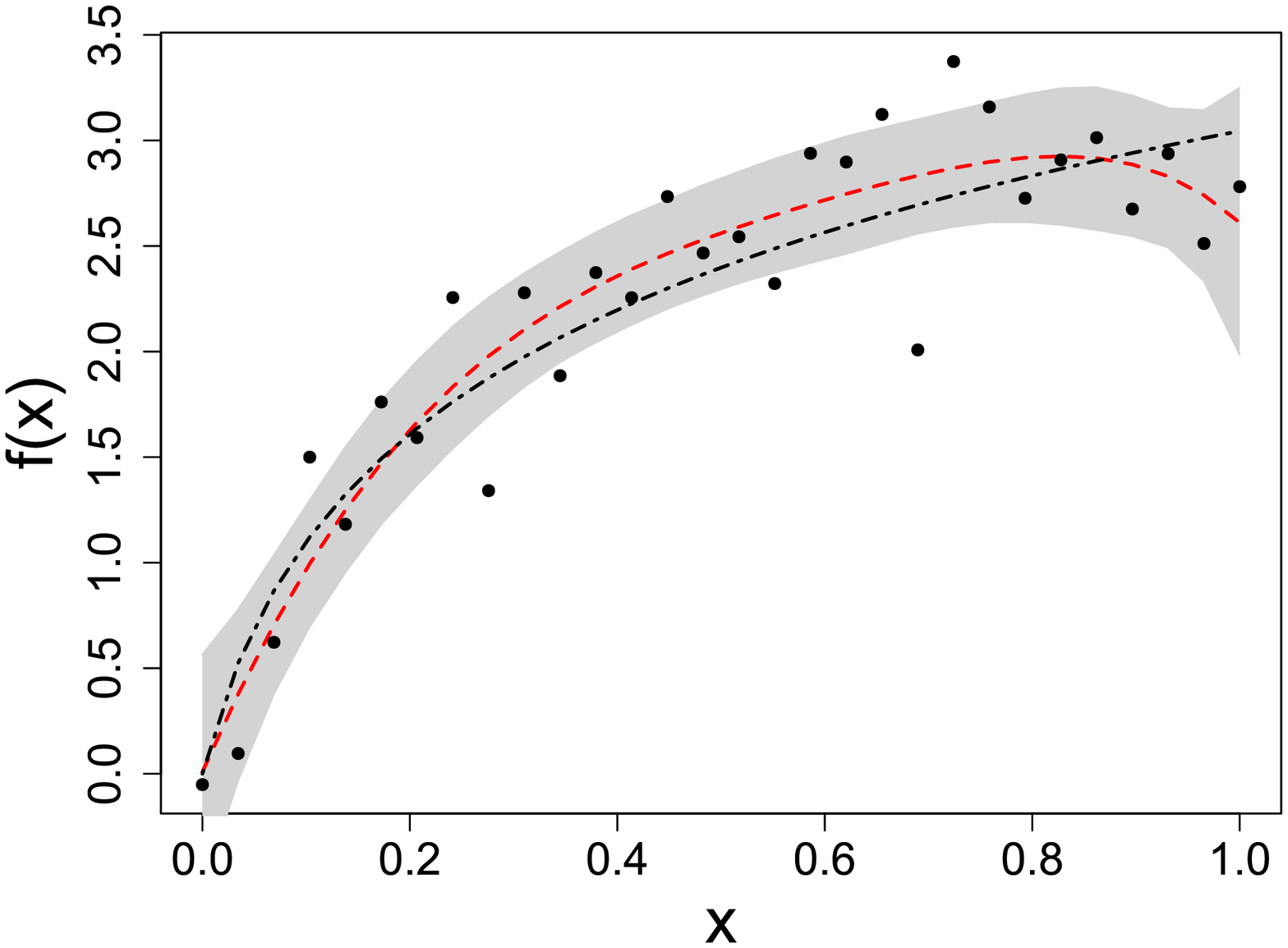}
      & \includegraphics[width=.3\textwidth]{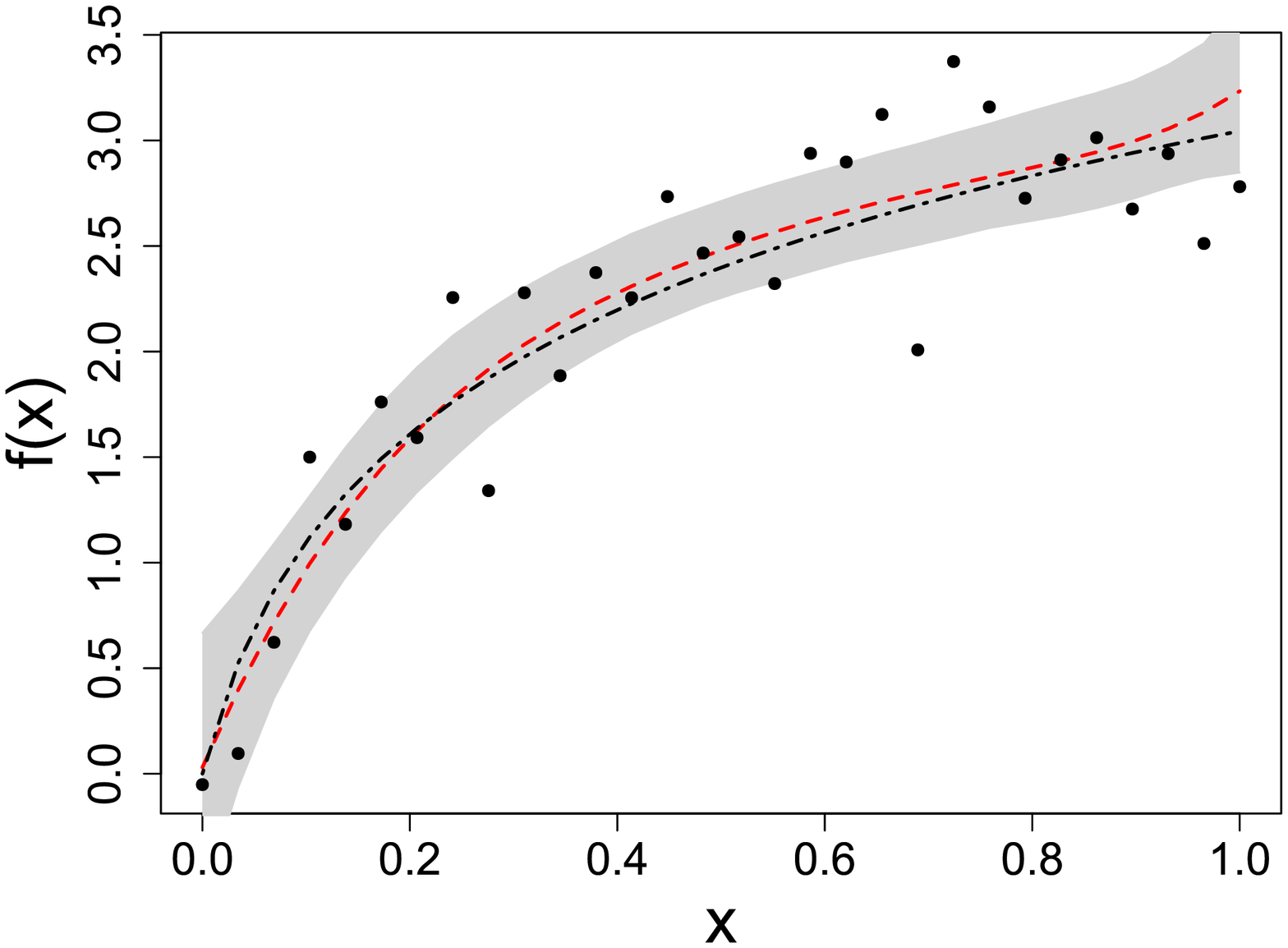}
 &\includegraphics[width=.3\textwidth]{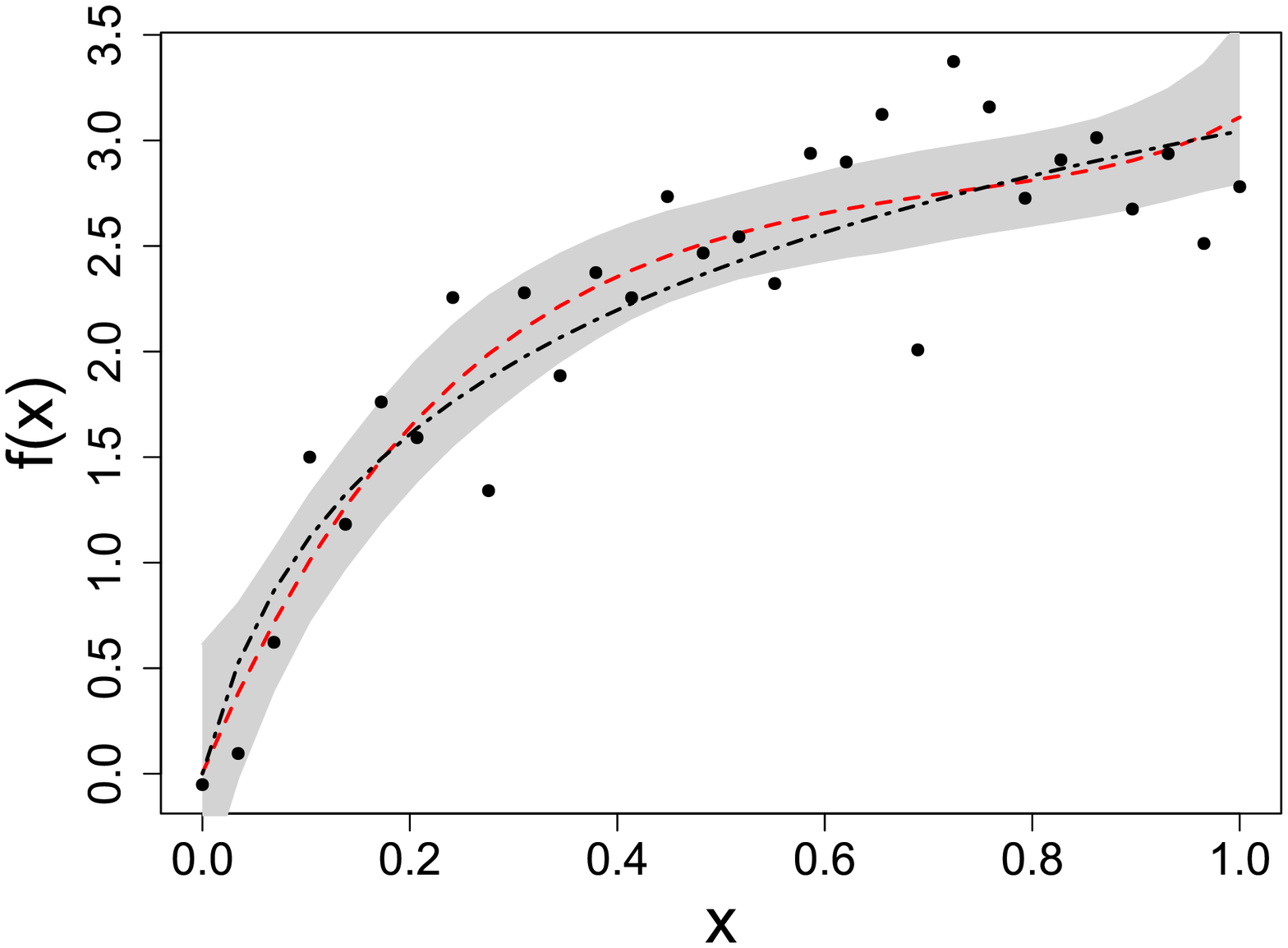}\\ 
    \rotatebox{90}{\hskip 25pt  logistic ($f_3$)} & \includegraphics[width=.3\textwidth]{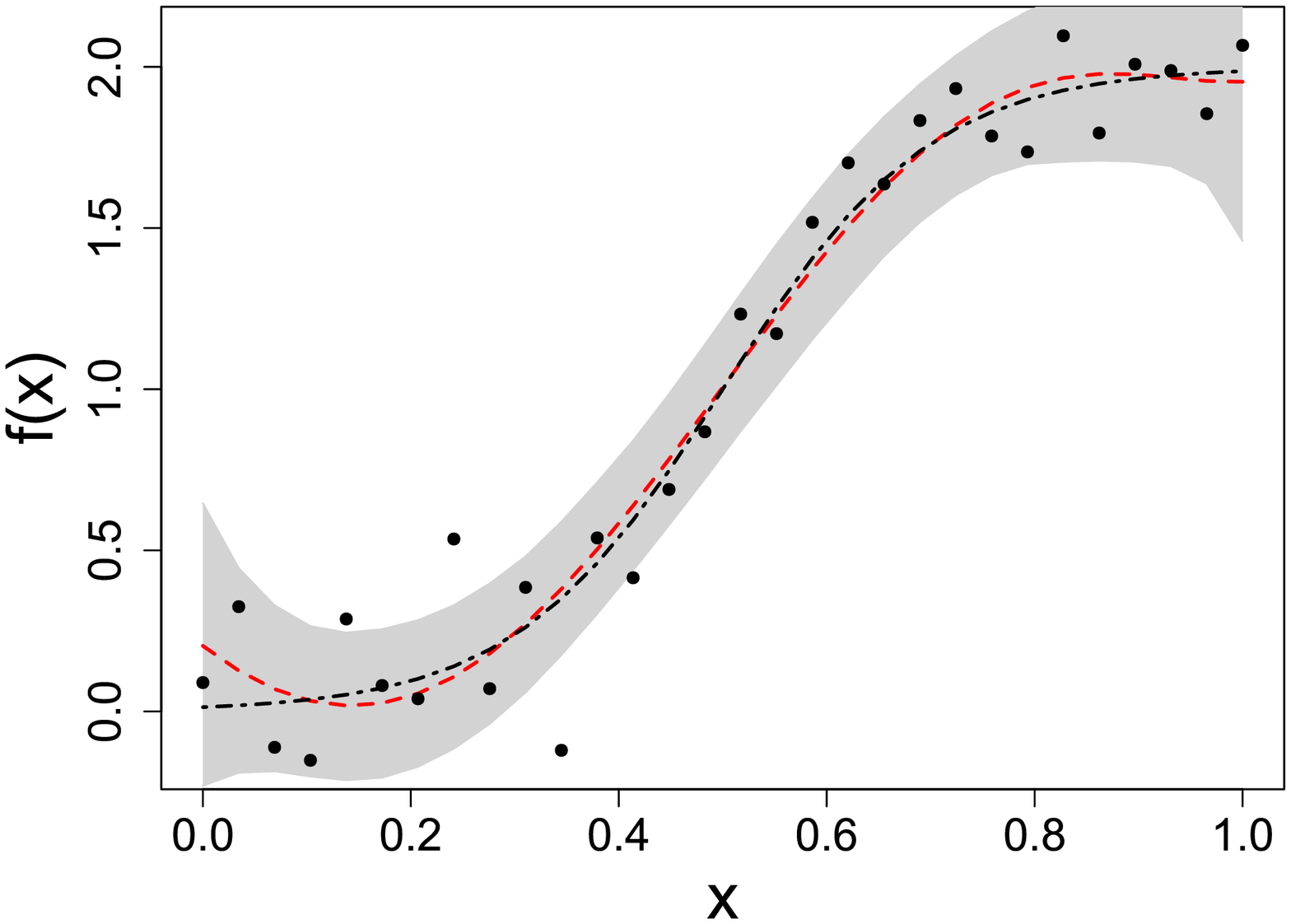}
      & \includegraphics[width=.3\textwidth]{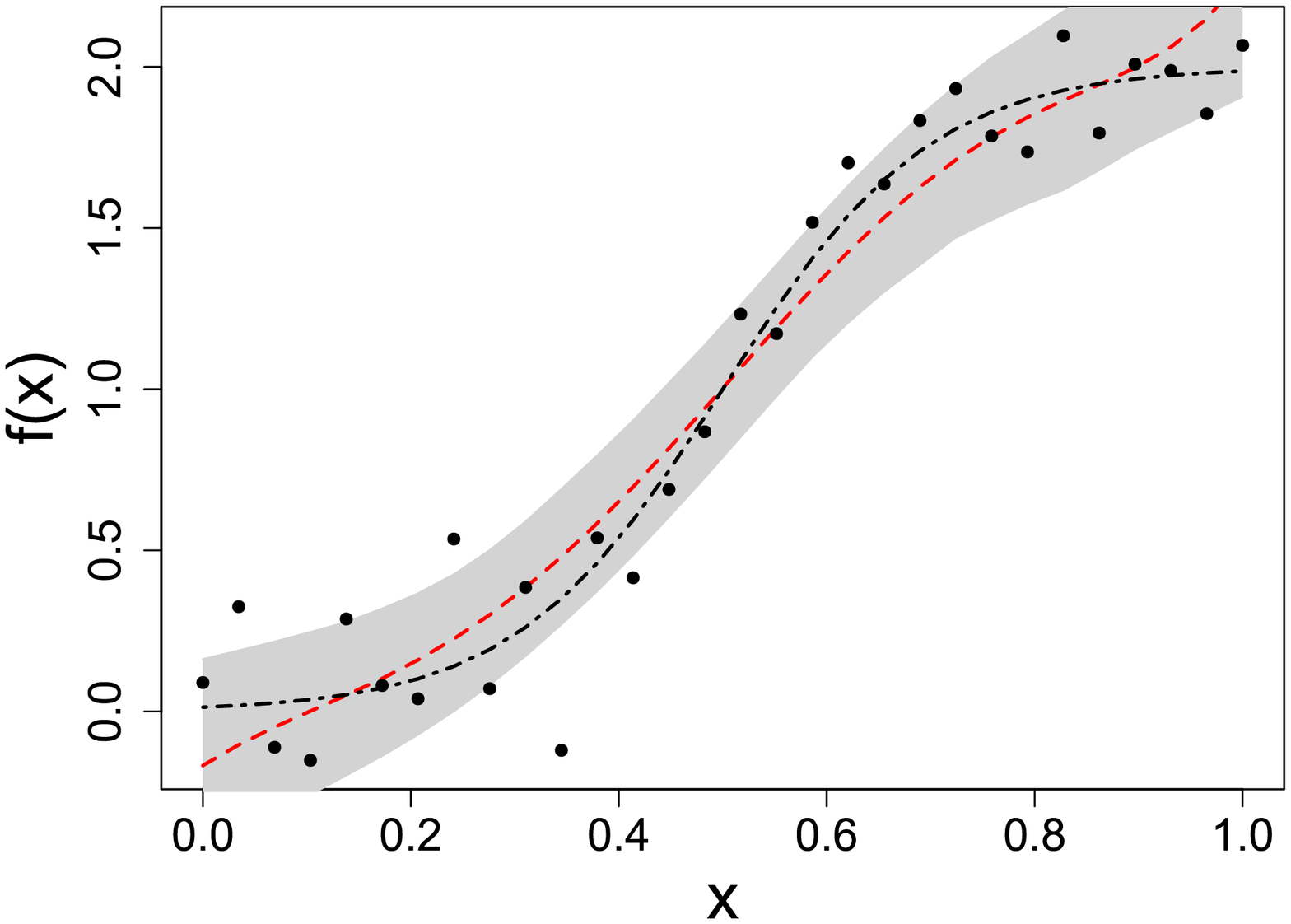}
 &\includegraphics[width=.3\textwidth]{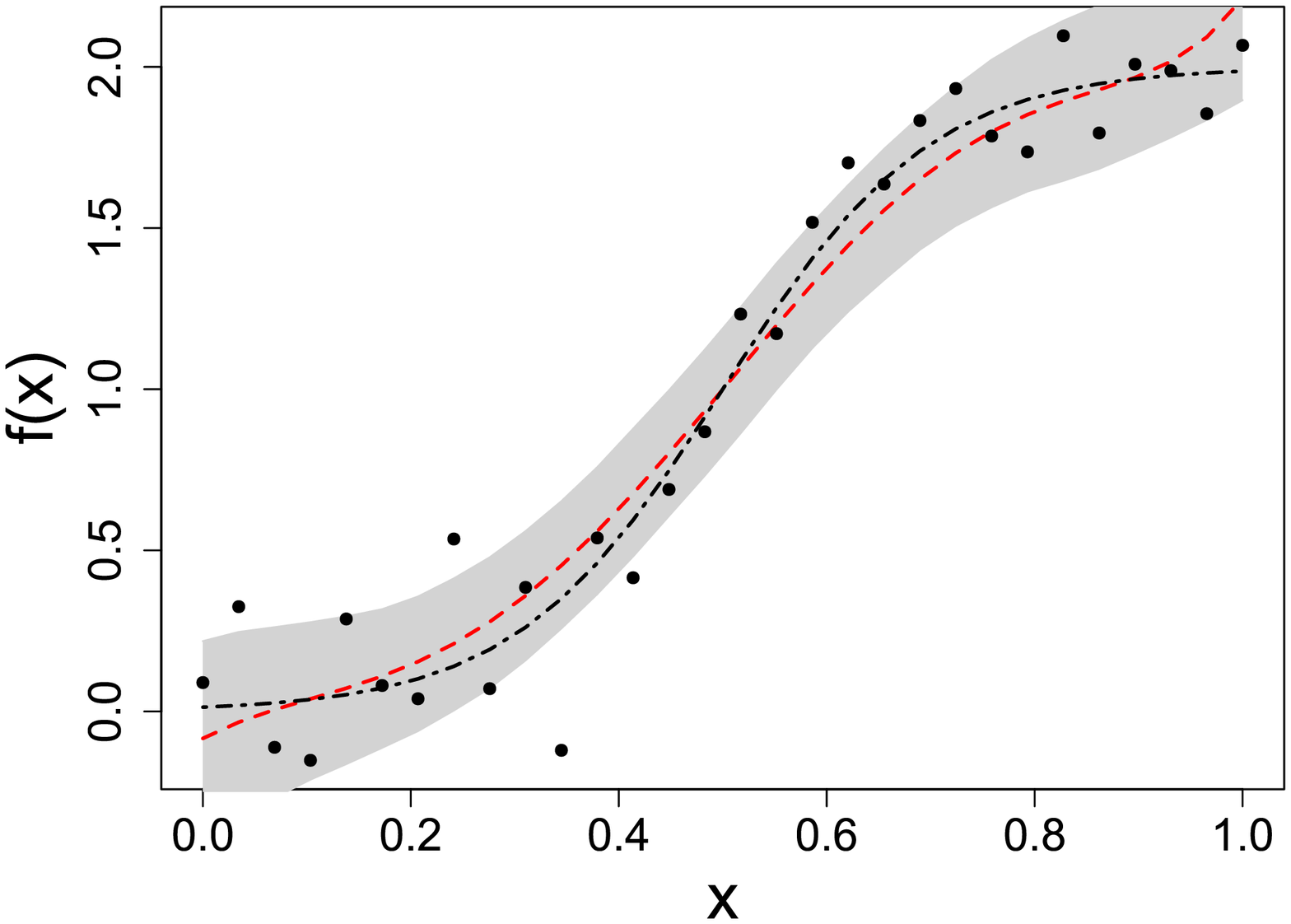}\\ 
  \end{tabular}
 \caption{Monotone polynomial regression fit and 95\% credible bands for noisy observations of monotone functions; the true functions (dash/dot black lines) are plotted together with the posterior mean of the polynomial fits (dashed red lines) for the three toy functions (rows) and three values of monotonicity parameter (columns)}\label{poly-reg}
\end{figure*}

	The monotone polynomial regression described above is fitted to data generated from the following monotone functions with additive normal noise at a grid of size $n=30$;
	\begin{align*}
	&f_1(x)=0.1+0.3x^3+0.5x^5+0.7x^7+0.9x^9,\\
	&f_2(x)=\log(20x+1),\\
	&f_3(x)=\frac{2}{1+\exp(-10x+5)}.
	\end{align*}
	Figure~\ref{poly-reg} shows the polynomial regression fits together with 95\% point-wise credible intervals at three steps of the SCMC with monotonicity parameters, $\tau=0$ (unconstrained polynomial regression), $\tau=1$, and $\tau=10^{5}$.

\subsection{Parameters Constrained to Measures on Manifolds}\label{sec:manifold}

Consider a problem posed on a popular statistics blog \footnote{\url{http://xianblog.wordpress.com/2014/03/24/mcmc-on-zero-measure-sets/}} of the bivariate standard Gaussian 
\begin{equation}\pi(X,Y) = {\cal N}\left(\left[\begin{array}{c}0\\0\end{array}\right],\left[\begin{array}{cc}1&0\\0&1\end{array}\right]\right)  I_\mathcal{A}(X,Y)\label{eq:xian}\end{equation}
where $\mathcal{A}=\{(X,Y): X^2-Y^2=1\}$ so that the resulting density lies on a pair of disjoint 1-dimensional manifolds embedded in the 2-dimensional space. 

The basic strategy for a Metropolis Hastings random walk algorithm is to propose $Y$ and define $X=\pm\sqrt{Y^2-1}$.  The acceptance ratio then requires the Jacobian for the change of measure to sample from the 1-dimensional manifold embedded within the 2-dimensional space.  Similar to Section \ref{scmc-toy} SCMC is applied by imposing a probabilistic constraint over a sequence of distributions admitting (\ref{eq:xian}) as its limit:
           \begin{equation*}
	\label{distance-metric-parabola}
	||(X,Y)||_\mathcal{A}^{\tau}:=2\Phi\left( -\tau_t\left|X^2-(Y^2+1)\right|  \right).
	\end{equation*}
Therefore, the sequence of distributions is defined by
	\begin{align}\label{eq:probit-manifold}
	\pi_t\left(X,Y\right) = &{\cal  N}\left(\left[\begin{array}{c}0\\0\end{array}\right],\left[\begin{array}{cc}1&0\\0&1\end{array}\right]\right) \\ \nonumber &2\Phi\left( -\tau_t\left|X^2-(Y^2+1)\right|  \right),
	\end{align}
Note that \[\lim_{\tau \rightarrow \infty} \Phi\left( -\tau\left|X^2-(Y^2+1)\right|  \right)  = \frac{1}{2}I_\mathcal{A}(X,Y)\] and consequently the factor of $2$ in (\ref{eq:probit-manifold}) is a normalizing constant for the indicator.  Although the SCMC samples are from (\ref{eq:xian}) with probability zero \[\int_X\int_Y\Big[\pi_t(x,y)-\pi\left(x,y\right)\Big]^2dxdy\rightarrow 0\mbox{ \ as \ }\tau_t\rightarrow\infty.\]
Consequently samples can be made arbitrarily close to the target without requiring the use of a transformation of variables or the use of a Jacobian when applying SCMC to sample from a density embedded within a lower dimensional manifold.   

\begin{figure*}[ht]
  \centering
 \includegraphics[width=\textwidth]{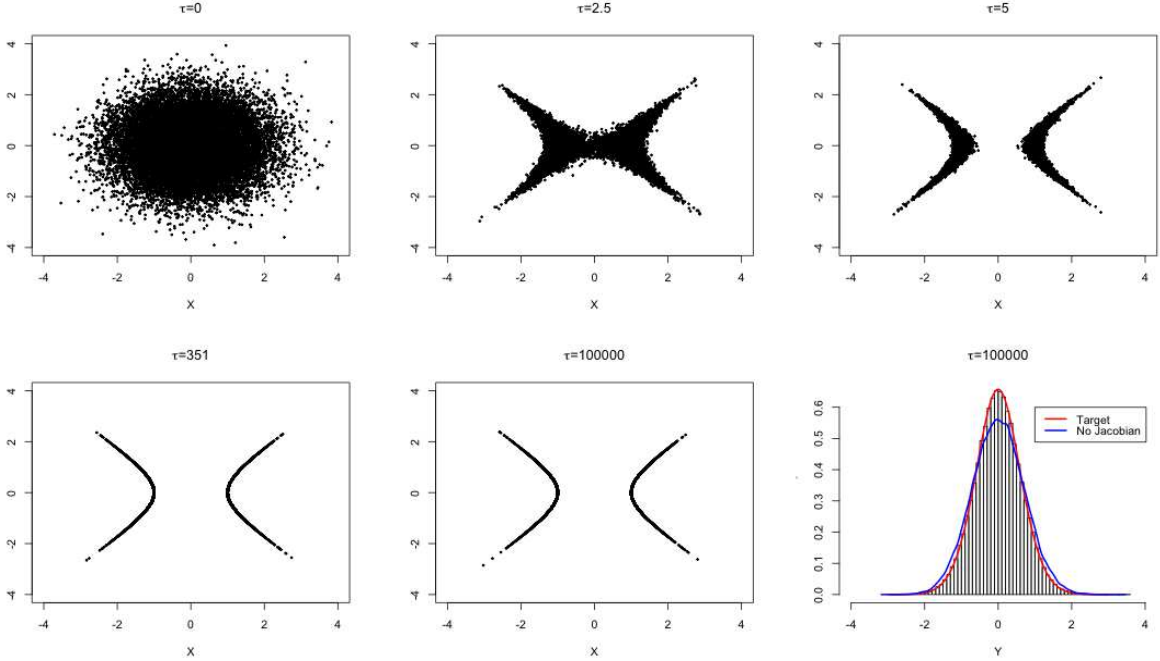} 
 \caption{Samples sequentially constrained to a parabolic manifold for different values of $\tau$ at the start and end of the sequence.  Bottom right: A Histogram of the $Y$ samples along with the true Target density (red) and the density when resulting from basic Metropolis Hastings without including the Jacobian term for sampling from the manifold.}\label{fig:parabola}
\end{figure*}

Starting from a sample of size 100,000 from unconstrained independent bivariate standard Gaussians (at $\tau=0$), a sequence of distributions with 1102 steps was defined with $\tau$ increasing up to 100,000.  At this final value of $\tau$ the probit function ensures that $P\left(|X^2-(Y^2+1)|>1.96/100000\right)=.05$.  Figure \ref{fig:parabola} shows samples from the constrained samples along with the SCMC and true target marginal distributions for $Y$.  Also included is the heavier tailed density one would achieve when using Metropolis Hastings while ignoring the Jacobian for the transformation to the lower dimensional manifold.

\section{Application Specific Constraints}\label{sec:app_constraints}

In some cases, the model structure can be exploited to define a sequence of small steps leading towards a fully enforced constraint.  This section showcases two such strategies.

\subsection{Differential Equation Models}\label{sec:ode}
	Ordinary Differential Equation (ODE) models implicitly describe the rate of change of system states $\mathbf{x}\left(\boldsymbol{\theta},\nu\right)$ with respect to time $\nu$ with the vector of parameters $\boldsymbol{\theta}\in\Theta$:
\begin{equation}
	\label{ODE}
	\frac{d\mathbf{x}\left(\boldsymbol{\theta},\nu\right)}{d\nu}=f\left(\mathbf{x}\left(\boldsymbol{\theta},\nu\right),\boldsymbol{\theta}\right).
	\end{equation}
	Because they are mechanistic models built from a theoretical understanding of a system, ODEs have the ability to model complex phenomena using relatively few but highly interpretable parameters.  The objective is to make inference about unknown parameters, $\boldsymbol{\theta}$, based on noisy observations,  $\mathbf{y}=\left(y_1,\ldots,y_n\right)^T$, that are available from the states (or a subset of them) at times, $\nu_1,\ldots,\nu_n$. Analytic solutions to (\ref{ODE}) often do not exist, consequently a numerical solver must be used to obtain  $\mathbf{x}\left(\boldsymbol{\theta},\nu\right)$ for a given initial condition $\mathbf{x}_0$.  When the initial condition is unknown, it is appended to the vector $\boldsymbol{\theta}$ with all other unknowns to estimate.  The posterior of $\boldsymbol{\theta}$ given the data is given by
	\begin{equation}
\label{ode-post}
	\pi\left(\boldsymbol{\theta}|\mathbf{y}\right)\propto \pi_0\left(\boldsymbol{\theta}\right) P\left(\mathbf{y}|\mathbf{x}\left(\boldsymbol{\theta},\nu\right),\boldsymbol{\theta}\right),
	\end{equation}
	where $\pi_0\left(\boldsymbol{\theta}\right)$ is the prior and the likelihood,\\  $P\left(\mathbf{y}|\mathbf{x}\left(\boldsymbol{\theta},\nu\right),\boldsymbol{\theta}\right)$, is centered on $\mathbf{x}\left(\boldsymbol{\theta},\nu\right)$ \citep{GelmanBoisJiang1996}.  
	Small changes in parameters can lead to large changes in the dynamics described by the ODE.  Consequently, multi-modality, ripples, ridges, and flat sections are common topological features of an ODE based likelihood\citep{CampbellSteele2011, CalderheadGirolami}, resulting in high rejection rates in random walk MCMC sampling schemes.  Exacerbating the challenging topology, when parameters are a mixture of discrete and continuous variables, the high posterior density regions can become separated into multiple distant modes.	To overcome topological likelihood difficulties, several groups have proposed model relaxation methods  through the introduction of a smoothing based approximation of the solution $\hat{\mathbf{x}}\left(\boldsymbol{\theta},\nu,b,\xi\right)\approx \mathbf{x}\left(\boldsymbol{\theta},\nu\right)$ \citep{RamsayEtAl2007,CalderheadEtAl2009,Brunel2008,LiangWu2008,CampbellSteele2011}.  The smoothing parameters $b$ and $\xi$ could represent the number of spline knots and penalty parameter,  parameters related to kernel smoothing, or Gaussian process parameters depending on the smoothing method employed.  The resulting approximation replaces the likelihood in (\ref{ode-post}) with $P\left(\mathbf{y}|\hat{\mathbf{x}}\left(\boldsymbol{\theta},\nu,b,\xi\right),\boldsymbol{\theta}\right)$. 

	From a similar perspective, we consider dependence on $\mathbf{x}\left(\boldsymbol{\theta},\nu\right)$ to be a model constraint and define the SCMC sequence using $\hat{\mathbf{x}}\left(\boldsymbol{\theta},\nu,b,\xi\right)$ as a relaxation from the strict adherence to the model.  Here, the model relaxation is performed through the inclusion of an additive residual fitting term defined by a kernel smoother whose role in the model diminishes gradually to facilitate filtering of the particles towards the target (\ref{ode-post}). 

Specifically, we make the approximation in the likelihood of (\ref{ode-post})
	\begin{equation}
	\label{kernel-model}
	\mathbf{x}(\boldsymbol{\theta},\nu)\approx\hat{\mathbf{x}}\left(\boldsymbol{\theta},\nu,b,\xi\right)=\mathbf{x}\left(\boldsymbol{\theta},\nu\right)+\xi \mathbf{e}\left(\boldsymbol{\theta},\nu,b\right),
	\end{equation}
where $\mathbf{e}(\boldsymbol{\theta},\nu,b)= K\left(\frac{y(\nu)-\mathbf{x}(\boldsymbol{\theta},\nu)}{b}\right)$ is a model discrepancy term estimated by smoothing the residuals $y(\nu)-\mathbf{x}(\boldsymbol{\theta},\nu)$ with a Nadaraya-Watson kernel $K(\cdot)$ with bandwidth $b$.  The scalar weight $\xi$ controls the contribution of the kernel smooth to the model.

As $b\rightarrow 0$, $\hat{\mathbf{x}}\left(\boldsymbol{\theta},\nu,b,\xi\right)$ is a data interpolator able to completely absorb the discrepancy between $ \mathbf{x}\left(\boldsymbol{\theta},\nu\right)$ and $y(\nu)$ for $\boldsymbol{\theta}\in\Theta$.  In this case \\$P\left(\mathbf{y}|\hat{\mathbf{x}}\left(\boldsymbol{\theta},\nu,b,\xi\right),\boldsymbol{\theta}\right)= 1\ \forall \boldsymbol{\theta}\in\Theta$, eliminating any impact of the likelihood from the posterior.	  To sequentially enforce the model fidelity constraint, $b$ increases from a small value while holding $\xi$ fixed. As $b$ becomes large enough to make the Kernel effectively uniform over the range of $\nu$, the model (\ref{kernel-model}) reduces to the ODE solution plus a constant, $E$, i.e.,
	\begin{align*}
	\lim _{b\rightarrow \infty}\hat{\mathbf{x}}\left(\boldsymbol{\theta},\nu,b,\xi\right)&=\mathbf{x}\left(\boldsymbol{\theta},\nu\right)+\xi\lim _{b\rightarrow \infty}\mathbf{e}\left(\boldsymbol{\theta},\nu,b\right)\\
	&=\mathbf{x}\left(\boldsymbol{\theta},\nu\right)+\xi E.
	\end{align*}
The coefficient $\xi$ controls the impact of the additive discrepancy between the estimated states and the ODE solution;
	\begin{align*}
	\lim_{\xi\rightarrow 0}\lim _{b\rightarrow \infty}\hat{\mathbf{x}}\left(\boldsymbol{\theta},\nu,b,\xi\right)&=\mathbf{x}\left(\boldsymbol{\theta},\nu\right)+\lim_{\xi\rightarrow 0}\xi E\\
	&=\mathbf{x}\left(\boldsymbol{\theta},\nu\right).
	\end{align*}
	In this model, the SCMC algorithm defines a sequence of models initially corresponding to an increasing schedule over the bandwidth parameter, $b$, while $\xi$ is held fixed at 1.  The constraint enforcement continues with  a decreasing schedule over $\xi$, with $b$ held fixed at its large value. That is, the $t^{\text{th}}$ distribution in the sequence is given by
	\begin{equation}
	\pi_{t}\propto \pi_0\left(\boldsymbol{\theta}\right) P\left(\mathbf{y}|\hat{\mathbf{x}}\left(\boldsymbol{\theta},\nu,b_t,\xi_t\right),\boldsymbol{\theta}\right).\label{eq:approx}
	\end{equation}
	for
	\begin{equation*}
	0=b_0<b_1<\ldots<b_{t^*}=b_{t^*+1}=\ldots=b_T,
	\end{equation*}
	and
	\begin{equation*}
	1=\xi_0=\xi_1=\ldots=\xi_{t^*}>\xi_{t^*+1}>\ldots>\xi_T=0.
	\end{equation*}

\subsubsection{Epidemiological ODE Model}
	We illustrate SCMC using (\ref{eq:approx}) on a Susceptible-Infected-Removed (SIR) epidemiological model for deaths due to the black plague.  During the black plague epidemic of 1666, the village of Eyam, UK, chose to quarantine themselves to avoid spreading the disease to neighboring villages.  As the plague swept through Eyam, the grave digger kept records of all those who perished during the outbreak.  Restricting ourselves to the second outbreak of the plague from June 19, 1666 to November 1, 1666, we partition the population of fixed size $N=261$ at time $\nu$ into groups of Susceptible  $S(\nu)$, Infected, $I(\nu)$, and Removed, $R(\nu)$ \citep{Raggett1982,Campbell13}.  Because there is no recovery from the plague, $R(\nu)$ corresponds to the number of deaths up to time $\nu$.   The epidemiological model for the rates of change of states $S(\nu),I(\nu)$ and $R(\nu)$ is given by:
	\begin{eqnarray}
\label{SIR}
	  \frac{dS\left(\nu\right)}{dt}&=&-\beta S\left(\nu\right)I\left(\nu\right),\nonumber\\
	  \frac{dI\left(\nu\right)}{dt}&=&\beta S\left(\nu\right)I\left(\nu\right)-\alpha I\left(\nu\right),\\
	   \frac{dR\left(\nu\right)}{dt}&=&\alpha I\left(\nu\right)\nonumber
	\end{eqnarray}
where $\beta$ describes the plague transmissivity and $\alpha$ describes the rate of death once an individual is infected.  More complex models have been considered elsewhere including vectors for disease transmission \citep[See for example][]{Massad04}, but the limitations of the Eyam data set do not support parameter estimation for more complex models. 

At time $0$ the population only consists of susceptible and infectious individuals therefore we have, $R(0)=0$ and $S(0)=N-I(0)$. Consequently $I_0=I(0)$ is included in the vector of parameters to estimate: $\boldsymbol{\theta}=(\alpha,\beta,I_0)$.

Based on a closed population of size $N$, the likelihood for the $n=136$ observed cumulative deaths $y(\nu)$ at times $\{\nu_1,\ldots,\nu_n\}$ are modeled as a binomial with expected value equal to the solution to $R(\boldsymbol{\theta},\nu)$ from (\ref{SIR}):
	\begin{align*}
	P\left(\mathbf{y}\mid R(\boldsymbol{\theta},\nu)\right)=\prod_{i=1}^{n} &{N \choose y_i} \left(\frac{R\left(\boldsymbol{\theta},\nu_i\right)}{N}\right)^{y_i}\\&\times\left(1-\frac{R\left(\boldsymbol{\theta},\nu_i\right)}{N}\right)^{\left(N-y_i\right)}.
	\end{align*}
   To evaluate the likelihood for $\theta$ using (\ref{kernel-model}) we first solve the ODE numerically to obtain $R(\boldsymbol{\theta},\nu)$, and then apply the kernel smoother to 
$y(\nu)-R\left(\boldsymbol{\theta},\nu\right)$.

	Prior distributions for $\alpha$ and $\beta$ were chosen to be $\text{gamma}\left(1,1\right)$. The prior distribution for $I_0$ was chosen to be a binomial$(N,5/N)$. Having both discrete and continuous parameters in the model compounds the difficulty of sampling from the posterior; MCMC can easily get trapped in local modes for $\alpha$ and $\beta$ corresponding to discrete values of $I_0$. The challenge of multi-modality is overcome by sequential enforcement of the model constraint. 
	
\begin{figure*}[ht]
        \centering
        \begin{subfigure}[b]{0.32\textwidth}
                \centering
                \includegraphics[width=\textwidth]{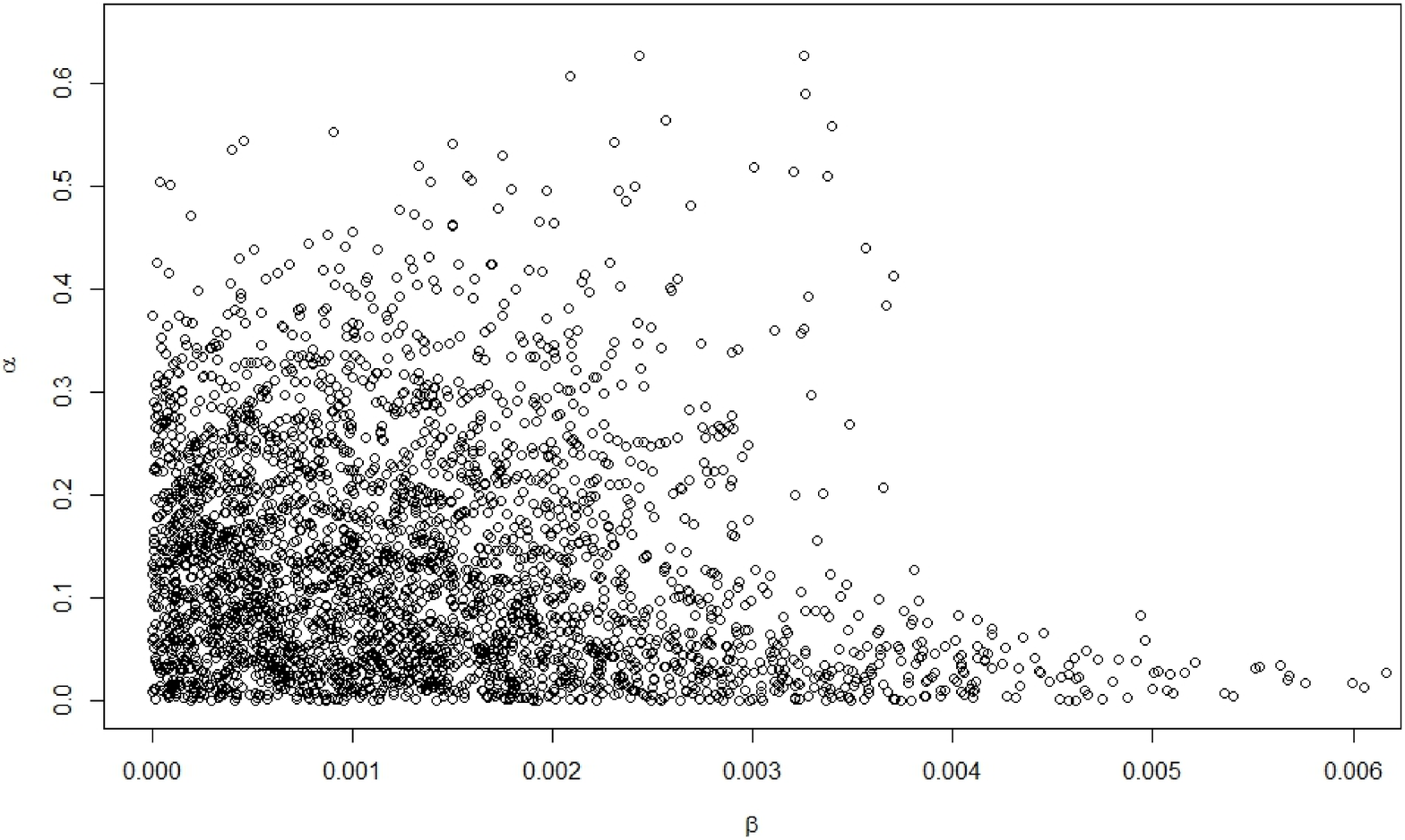}
                \caption{$b=2$, $\xi=1$}
                \label{SIR1}
        \end{subfigure}
        \begin{subfigure}[b]{0.32\textwidth}
                \centering
                \includegraphics[width=\textwidth]{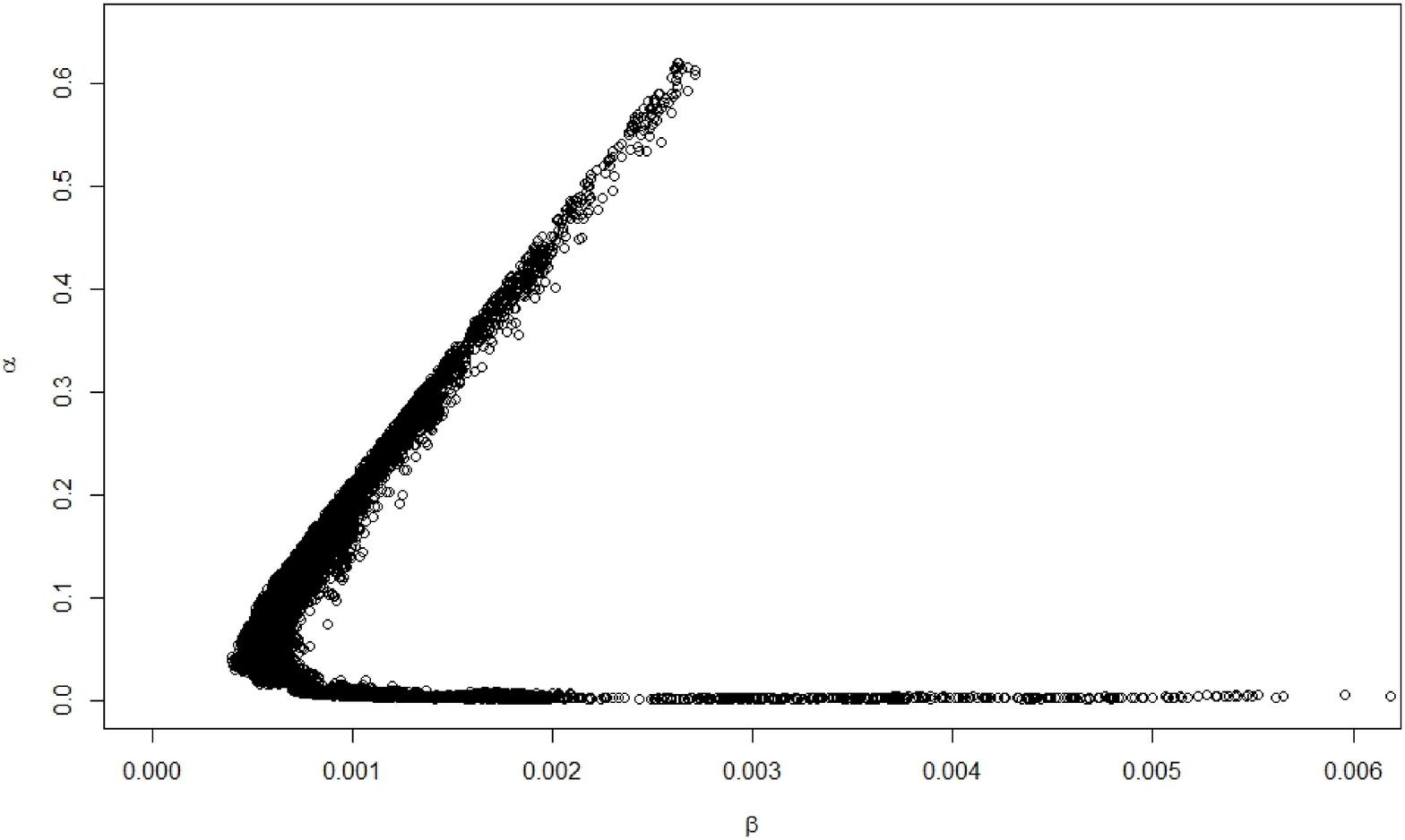}
                \caption{$b=12$, $\xi=1$}
                \label{SIR2}
        \end{subfigure}
        \begin{subfigure}[b]{0.32\textwidth}
                \centering
                \includegraphics[width=\textwidth]{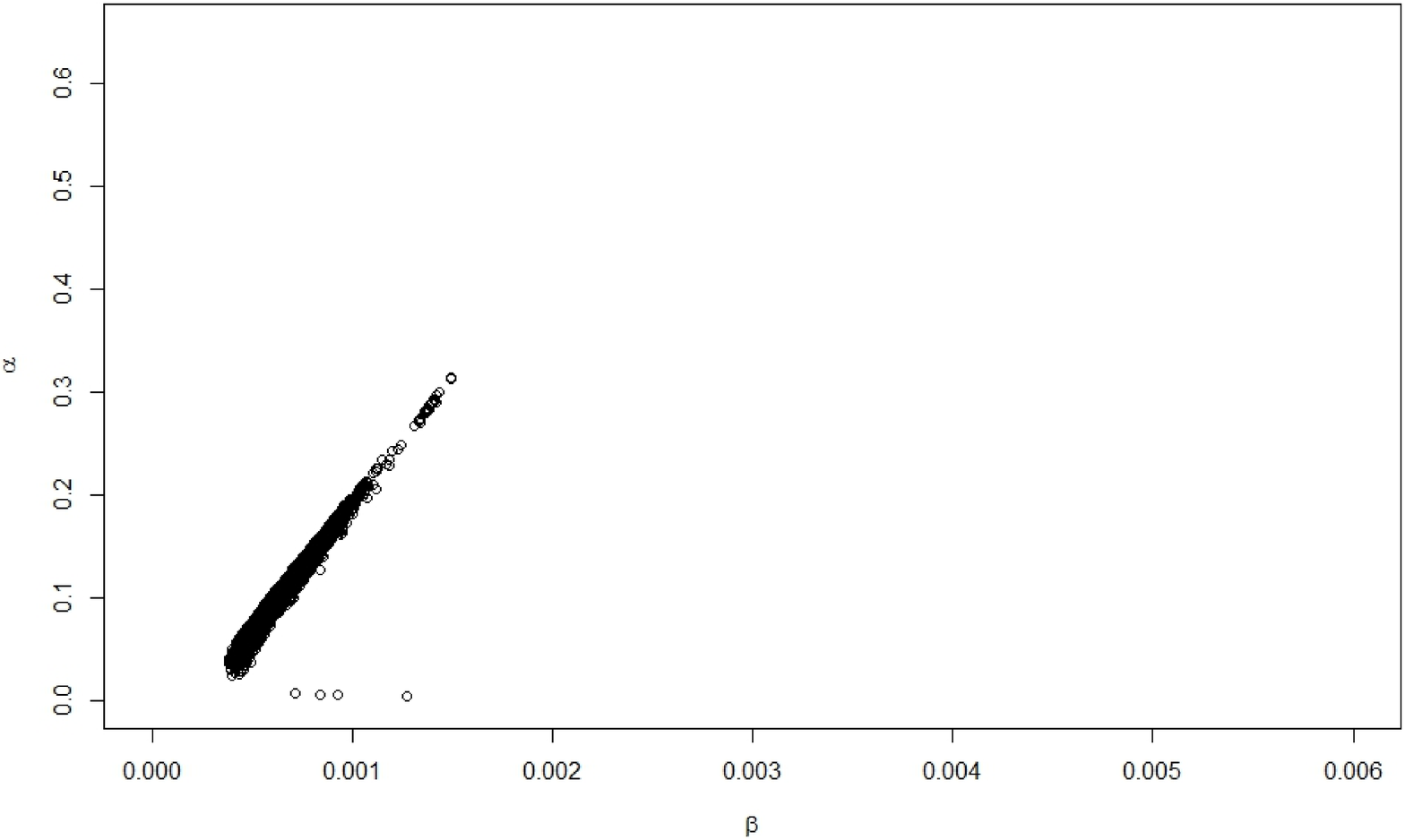}
               \caption{$b=26$, $\xi=1$}
                \label{SIR3}
        \end{subfigure}
        \begin{subfigure}[b]{0.32\textwidth}
                \centering
                \includegraphics[width=\textwidth]{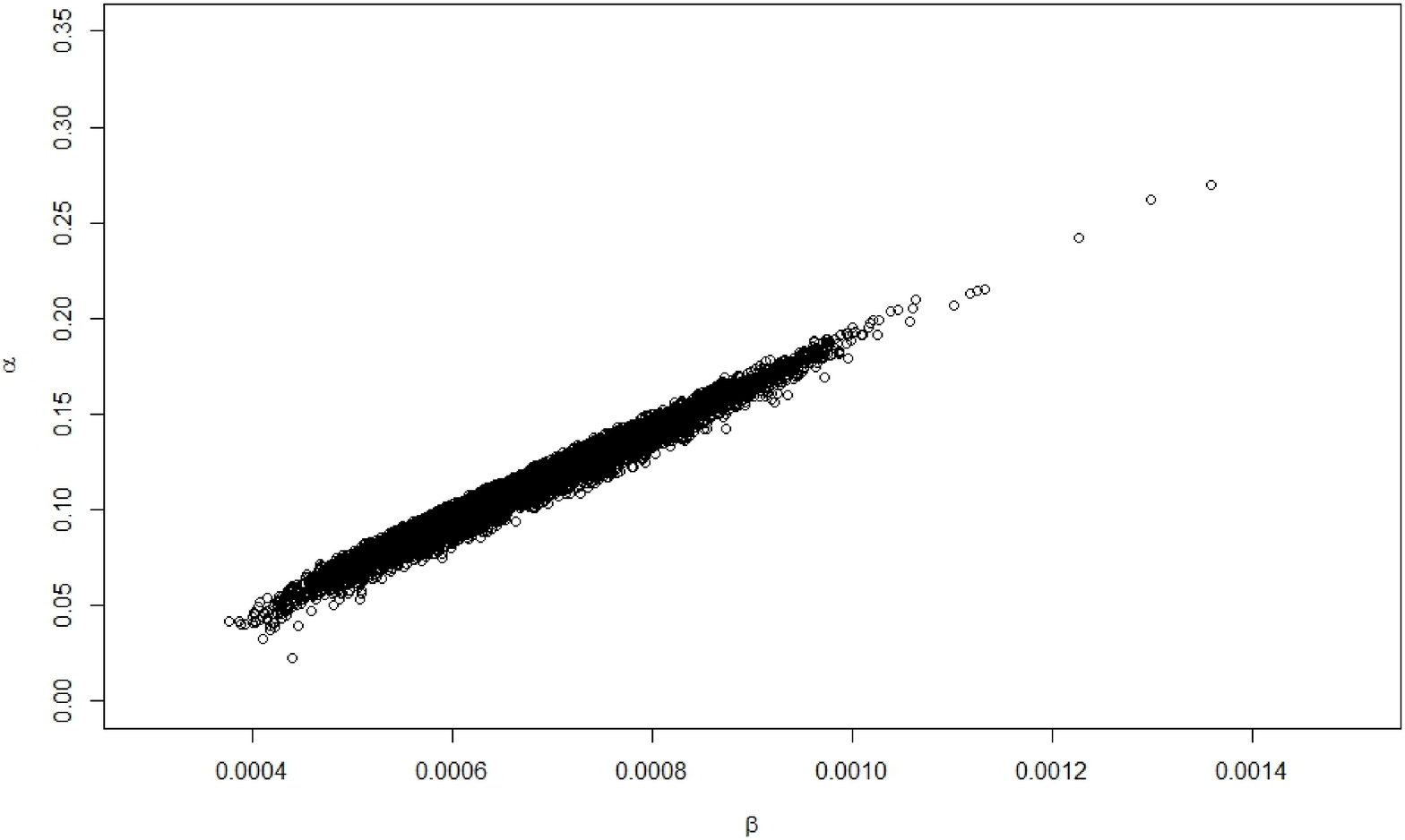}
               \caption{$b=26$, $\xi=0.95$}
                \label{SIR4}
        \end{subfigure}
        \begin{subfigure}[b]{0.32\textwidth}
                \centering
                \includegraphics[width=\textwidth]{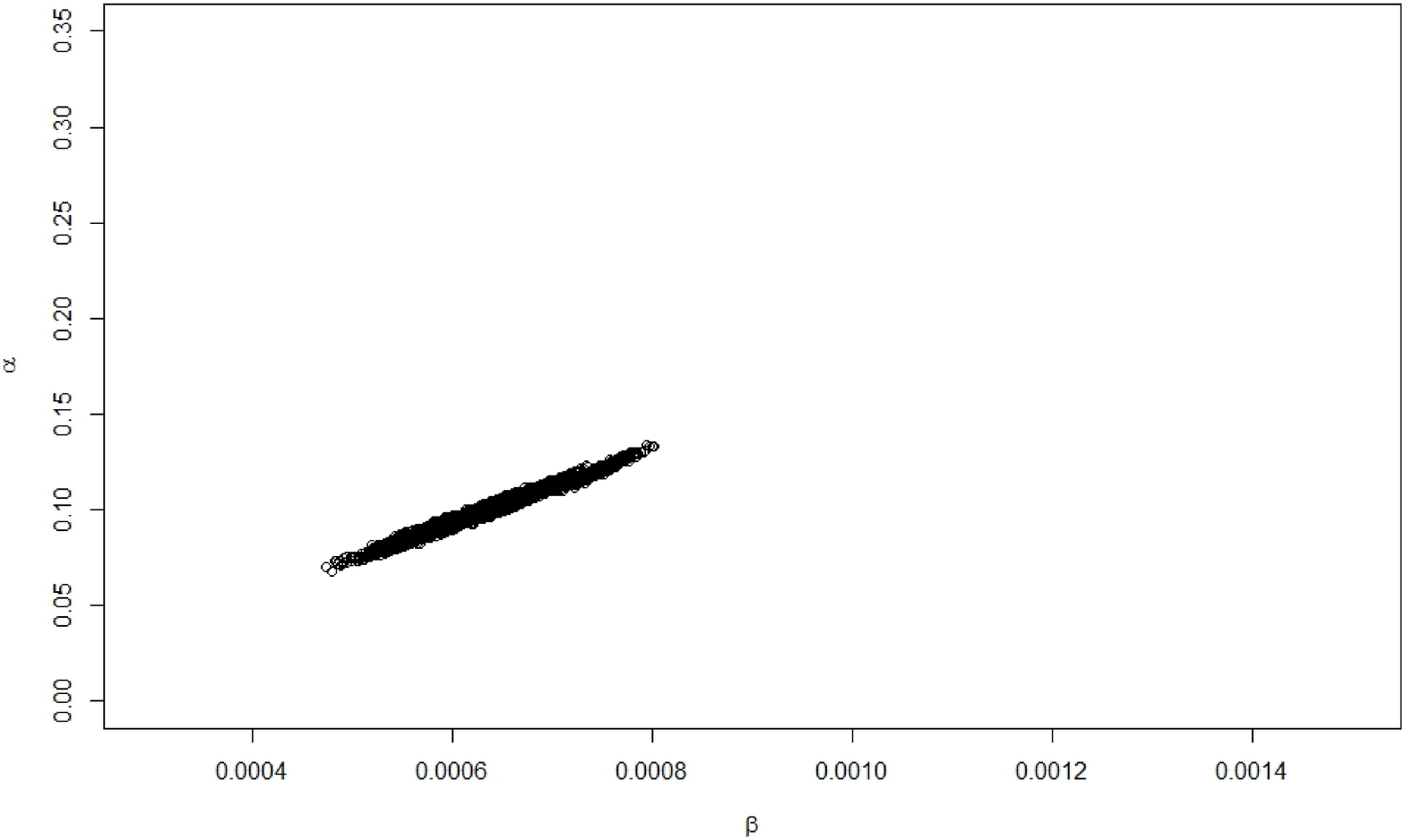}
               \caption{$b=26$, $\xi=0.5$}
                \label{SIR5}
        \end{subfigure}
      \begin{subfigure}[b]{0.32\textwidth}
                \centering
                \includegraphics[width=\textwidth]{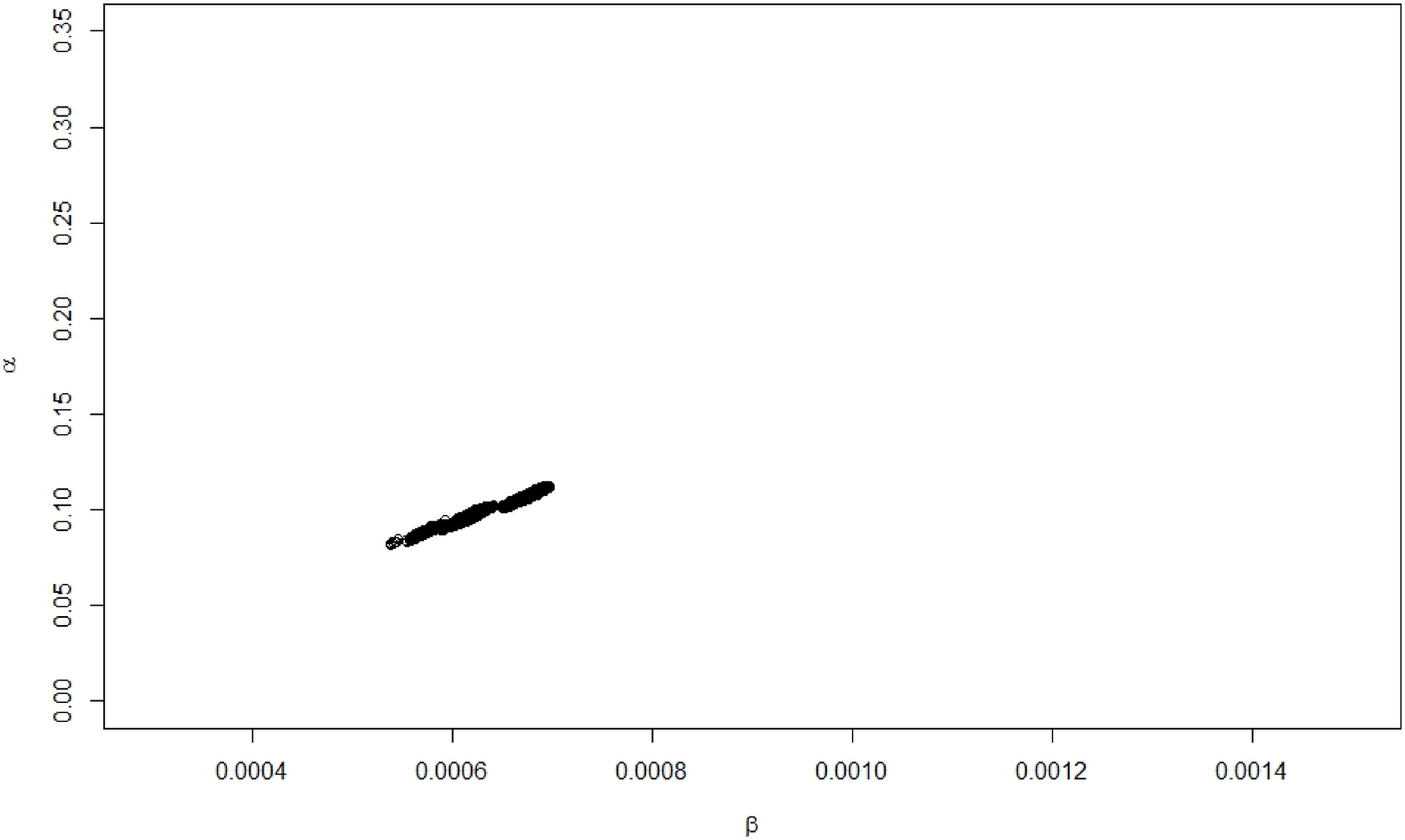}
               \caption{$b=26$, $\xi=0$}
                \label{SIR6}
        \end{subfigure}
     
        \caption{The SIR model- evolution of the posterior as a result of increasing the bandwidth, $b$, (a-c), and decreasing the coefficient, $\xi$, (d-f). }\label{SIR-post}
\end{figure*}

	Figures~\ref{SIR1}, \ref{SIR2} and \ref{SIR3} present the marginal posterior of (\ref{eq:approx}) for $\alpha$ and $\beta$	for three increasing values of  $b$, while holding $\xi=1$ fixed. The parameter densities are distributed in the shape of a boomerang in Figure~\ref{SIR2}. The lower part of the boomerang refers to parameter values whose corresponding states are non-smooth and deviate from the ODE solution in such a way that can be sufficiently accommodated by the smoother. By increasing $b$, the smoother reduces to a constant and the corresponding particles are filtered out.

\begin{figure}
\centering
\includegraphics[width=.48\textwidth]{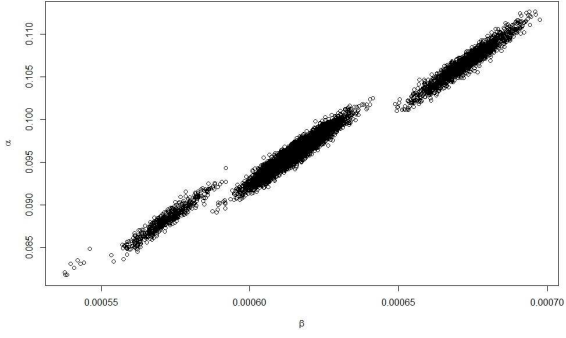}
               \caption{The SIR model - joint posterior distribution of the model parameters for $b=26$ and $\xi=0$. The three large clouds of particles correspond to $I_0=6$, $I_0=5$ and $I_0=4$, respectively, from left to right.}
                \label{finalode}
\end{figure}

\begin{figure}[h!]
\centering
\includegraphics[width=.48\textwidth]{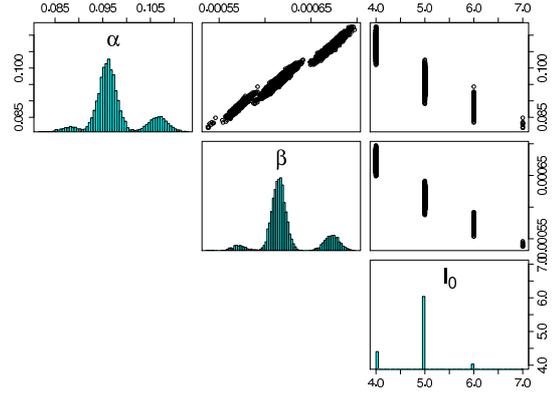}
               \caption{The SIR model - marginal (diagonal) and bivariate joint (off-diagonal) posterior distribution of the model parameters and initial states for $b=26$ and $\xi=0$. }
                \label{mat-plot}
\end{figure}

Figures~\ref{SIR4}, \ref{SIR5} and \ref{SIR6} show the joint posterior samples holding $b=26$ fixed and decreasing $\xi\rightarrow 0$. Figure~\ref{finalode} shows the posterior sample for the final step of the algorithm based on (\ref{ode-post}) after fully enforcing the model and rescaling axes for better visualization. The three large clouds of parameter values represent the posterior modes that refer to $I_0=6$, $I_0=5$ and $I_0=4$ from left to right, respectively. The marginal and bivariate joint posterior distributions are illustrated in Figure~\ref{mat-plot} highlighting the multi-modality  and topological challenges of the model.

\begin{figure*}[ht]
\centering
\begin{subfigure}[b]{0.45\textwidth}
                \centering
                \includegraphics[width=\textwidth]{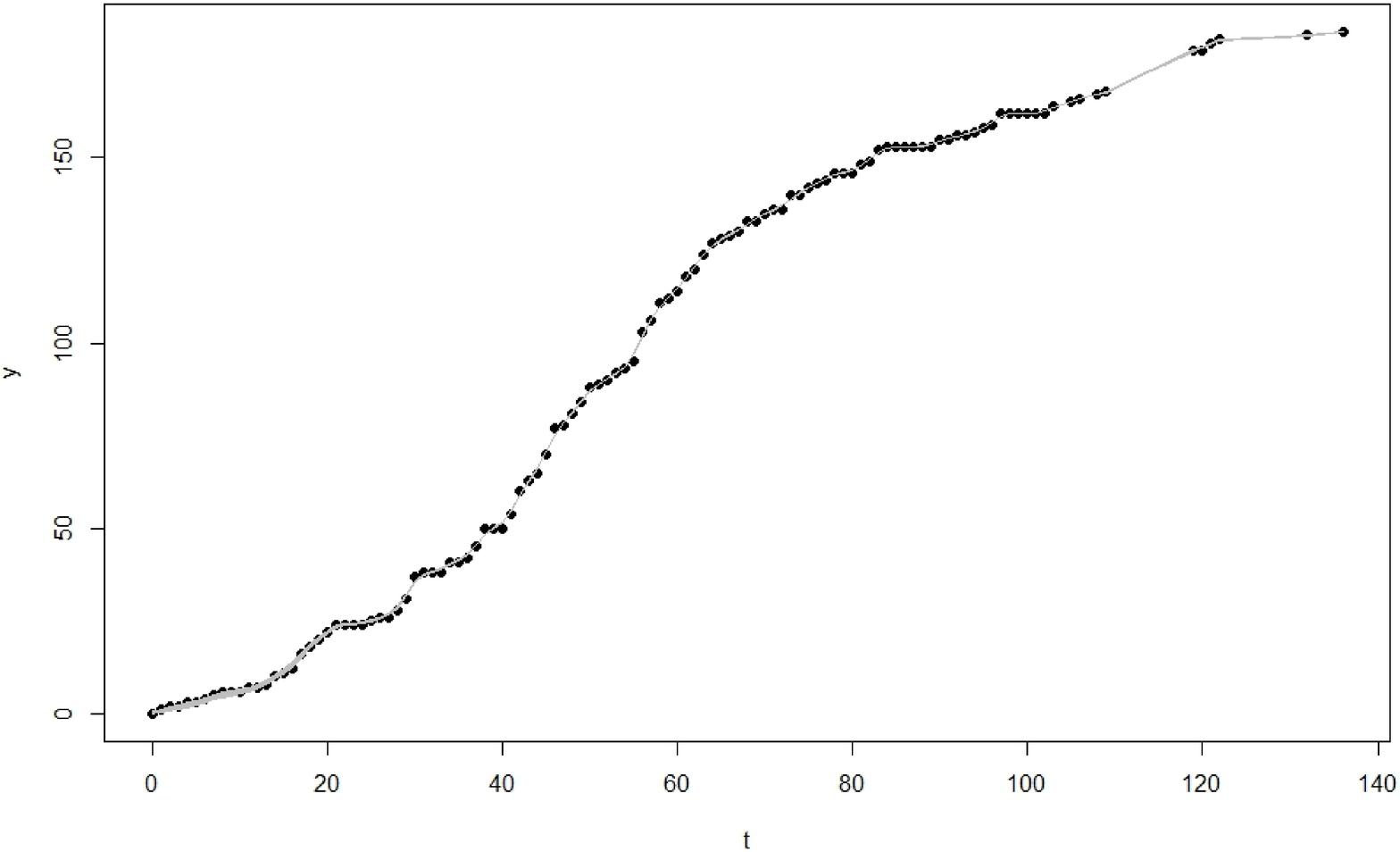}
               \caption{$b=2, \xi=1$}
                \label{sp-a}
        \end{subfigure}\\
 \begin{subfigure}[b]{0.45\textwidth}
                \centering
                \includegraphics[width=\textwidth]{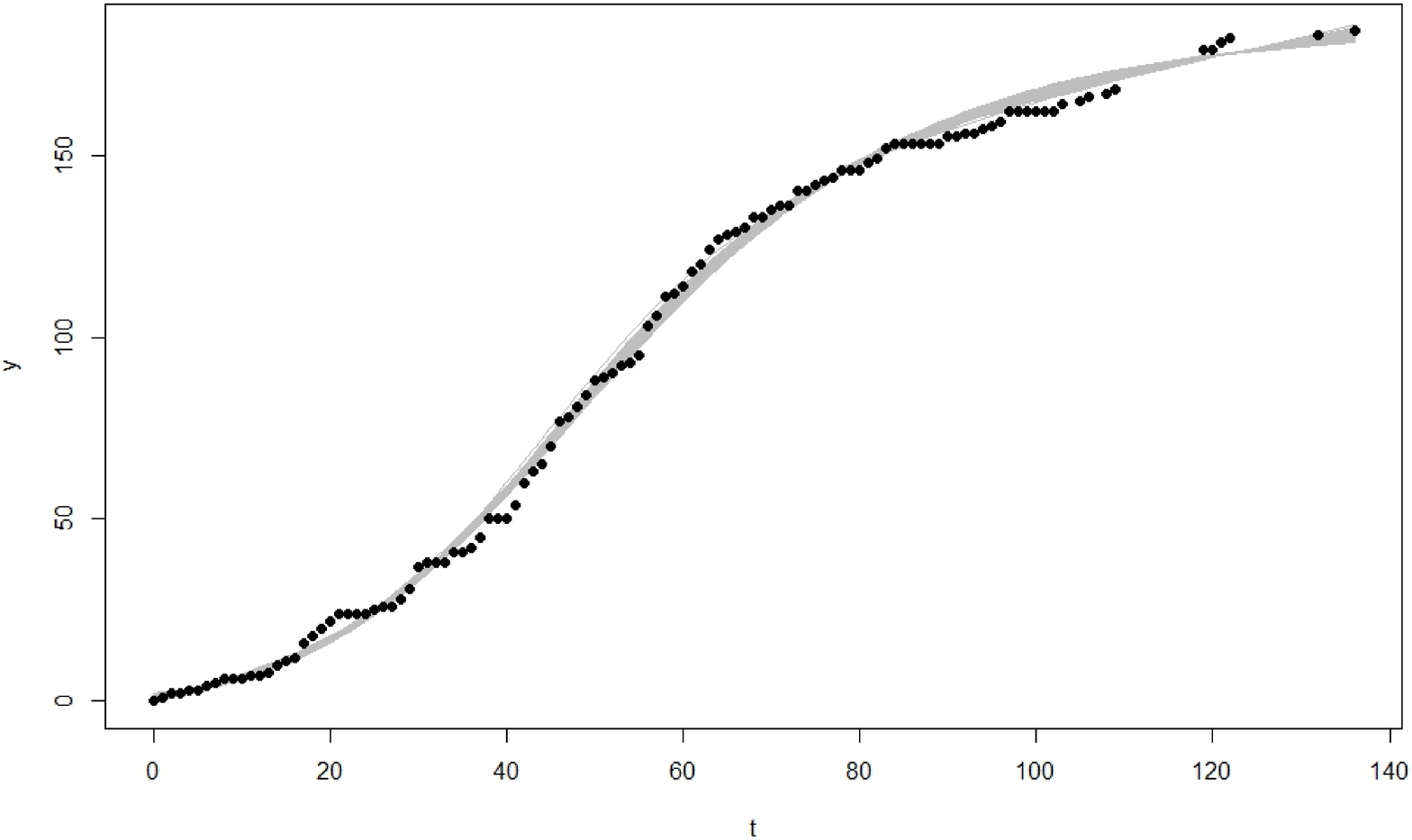}
               \caption{$b=26, \xi=1$}
                \label{sp-a}
        \end{subfigure}
  \begin{subfigure}[b]{0.45\textwidth}
                \centering
                \includegraphics[width=\textwidth]{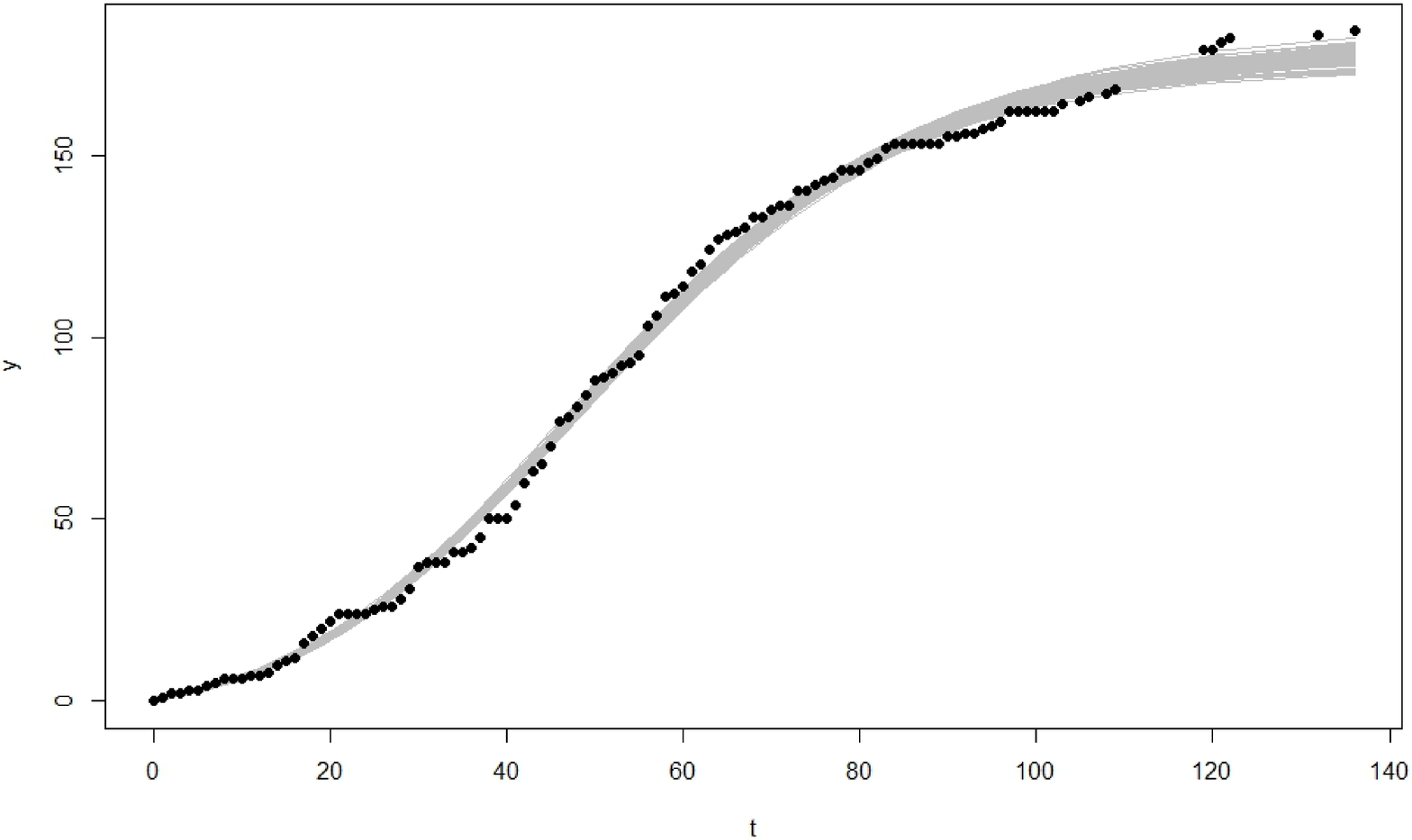}
               \caption{$b=26,\xi=0$}
                \label{sp-b}
        \end{subfigure}
\caption{100 posterior sample paths plotted against the data for (a) $b=2, \xi=1$, \\ (b) $b=26, \xi=1$ and (c) $b=26,\xi=0$.}\label{fits-SIR}
\end{figure*}

A sample of 100 fits to the data is given in Figure~\ref{fits-SIR} based on the posterior at early ($b=2$ and $\xi=1$), intermediate ($b=26$ and $\xi=1$), and final ($b=26$ and $\xi=0$) steps of filtering schedule.  Although the fits to the data are quite similar in the panels of  Figure~\ref{fits-SIR}, they correspond to extreme differences in $\pi_t(\theta\mid y)$ as seen in Figure~\ref{SIR-post}.

\subsection{Approximate Bayesian Computation}\label{sec:abc}
	
	Approximate Bayesian Computation (ABC)  is used for inference in cases where the likelihood is intractable or expensive to evaluate but simulated data can be easily generated from the model \citep{Tavare97} . For a given set of parameters $\boldsymbol{\theta}$, simulated data, $\mathbf{z}$, is generated  and an approximation to the posterior is constructed by replacing the likelihood with an indicator function $I_{\mathcal{A}}\left(\mathbf{z}\right)$ over the matching set, $\mathcal{A}$, comparing $\mathbf{y}$ and the $\boldsymbol{\theta}$ generated $\mathbf{z}$ in the ABC posterior:
	\begin{equation}\label{eq:abc_posterior}
	\pi_A\left(\boldsymbol{\theta},\mathbf{z}|\mathbf{y}\right)=\frac{\pi\left(\boldsymbol{\theta}\right)P\left(\mathbf{z}|\boldsymbol{\theta}\right)I_{\mathcal{A}}\left(\mathbf{z}\right)}{\int_{A}\pi\left(\boldsymbol{\theta}\right)P\left(\mathbf{z}\mid \boldsymbol{\theta}\right)d\mathbf{z}}.
	\end{equation}

	The matching criteria defining $\mathcal{A}$ is a measure of similarity between the summary statistics, $\boldsymbol{s}(\cdot)$, of the simulated and observed data.  The approximation error of ABC depends on this matching criteria and is reduced to zero if $\boldsymbol{s}(\cdot)$ is sufficient for $\boldsymbol{\theta}$ and $\mathcal{A}$ is defined based on an exact match, i.e.
	\begin{equation*}
	\mathcal{A} = \left\{\mathbf{z} \mid \boldsymbol{s}\left(\mathbf{z}\right)=\boldsymbol{s}\left(\mathbf{y}\right)\right\}.
	\end{equation*}
	
For continuous data $P(\mathbf{z}\in \mathcal{A})=0$, consequently, a distance metric $\rho(\cdot,\cdot)$ and tolerance level $\epsilon>0$ are introduced to reduce the ABC approximation error giving  the matching criteria:
	\begin{equation}\label{A}
	\mathcal{A} = \left\{\mathbf{z} \mid\rho\left(\boldsymbol{s}\left(\mathbf{z}\right),\boldsymbol{s}\left(\mathbf{y}\right)\right)<\epsilon\right\}. 
	\end{equation}
	In practice, (even for discrete data where an exact match is possible) non-trivial sufficient statistics are rarely known. Instead,  an ``approximately sufficient" set is used for $\boldsymbol{s}(\cdot)$.  Selection of a most informative subset of summary statistics has been explored in the literature; \cite{Joyce08} consider sequential addition of summary statistics where the effectiveness of addition of a new summary statistic is tested using a likelihood ratio test. Some drawbacks of their method are discussed in \cite{Marin12}. \cite{Ratmann09} considers a similar approach; they assess the sensitivity of summaries to change in the model parameters by measuring the derivative of their expectations with respect to the corresponding parameters. Summaries with smaller variance and higher sensitivity to parameter changes are preferable choices as they are more informative and hence bring $\boldsymbol{s}(\cdot)$ closer to sufficient statistics.

	 When a collection of candidate summary statistics are available for $\boldsymbol{s}(\cdot)$, a safe strategy to reduce the approximation error is to use the entire collection in (\ref{A}), since including more summary statistics should make $\boldsymbol{s}(\cdot)$ closer to the sufficient statistics.  However this comes with a computational trade off and makes the matching criteria more difficult to apply.

In the simplest version of the ABC algorithm, parameter values, $\boldsymbol{\theta}^*$, are generated from the prior.  If the associated simulated data $\mathbf{z}\in \mathcal{A}$, $\boldsymbol{\theta}^*$ is accepted as a sample value from the posterior and it is otherwise rejected. Note that this can be deeply inefficient for diffuse prior distributions.  SMC ABC algorithms define bridging distributions of \eqref{eq:abc_posterior} \\through a decreasing sequence of $\boldsymbol{\epsilon}$ \citep[see for example, ][]{Sisson09, Beaumont09, Moral09}.  While decreasing $\epsilon$ is one way of increasing the strictness of a model constraint and therefore fits into the SCMC framework, in this section we showcase the flexibility of using a constraint based algorithm by developing an alternative strategy.  Here, an alternative definition of sequentially increasing the model adherence constraint is developed through the iterative addition of summary statistics to the matching criteria.  

Suppose that $\boldsymbol{s}(\cdot)=\left(s_1(\cdot),\ldots,s_T(\cdot)\right)^T$, is the collection of available summary statistics. The target posterior is one that corresponds to a matching criteria that uses the entire collection of $T$ summary statistics.  An approximate posterior with a smaller set of summary statistics is likely to be more diffuse over the parameter space and therefore easier to sample.   The sequence of approximate posterior distributions, $\{\pi_{A_{t}}\}_{t=1}^T$, is defined based on a decreasing sequence of acceptance sets,
	\begin{equation*}
	\mathcal{A}_1\supseteq \mathcal{A}_2\supseteq \ldots \supseteq \mathcal{A}_T,
	\end{equation*}
	where
	\begin{align*}
	\mathcal{A}_{t}=\{\mathbf{z}|&\rho\left(s_1\left(\mathbf{z}\right),s_1\left(\mathbf{y}\right)\right)<\epsilon_1,\\&\ldots,\rho\left(s_{t}\left(\mathbf{z}\right),s_{t}\left(\mathbf{y}\right)\right)<\epsilon_{t} \}.
	\end{align*}
	The constraint parameter in this case is the number of summary statistics included up to stage $t\in\{0,1,.\ldots,T\}$. The sequence of posteriors defined by adherence to additional summary statistics results in sequentially constraining the posterior.  At each iteration the bridging distribution will move $\boldsymbol{\theta}$ closer to the target posterior.  The exception to this sequential constraining is if the newly included summary statistic is based on redundant information already contained in the set of summaries, in which case the bridging distribution will remain unaffected at its inclusion. 

	Algorithm~\ref{scmc-abc} outlines the SCMC ABC algorithm that generates parameter values according to a sequence of approximate posterior distributions constructed as described above. In the following section,  Algorithm~\ref{scmc-abc} is used for parameter estimation from a chaotic dynamical model to illustrate the effectiveness of the algorithm. 
	
	\begin{algorithm}[t]
	\caption{SCMC ABC }\label{scmc-abc}
	\begin{algorithmic}[1]
	\renewcommand{\algorithmicrequire}{\textbf{Input:}}
	\renewcommand{\algorithmicensure}{\textbf{Return:}}
	\Require Sequence of matching criteria $\{A_{t}\}_{t=1}^T$
	
	MCMC transition kernels $K_{t}$.\\
	
	Generate a sample from $\pi_{A_{1}}\left(\boldsymbol{\theta},\mathbf{Z}|\mathbf{y}\right)$:
	\begin{itemize}
	\renewcommand{\labelitemi}{ }
	\item $i\gets 0$
	\While{$i< N$} 
	
	generate $\boldsymbol{\theta} \sim \pi\left(\boldsymbol{\theta}\right)$
	
	generate $\mathbf{Z}=\left(\mathbf{z}_1,\ldots, \mathbf{z}_M\right)$
	
	$w \gets \sum_{k=1}^{M}\mathbb{I}_{A_1}\left(\mathbf{z}_k\right)$
	
	\If {$w>0$} 
	
	\hskip 30pt $i\gets i+1$
	
	\hskip 30pt $\left(\boldsymbol{\theta}_{1}^{\left(i\right)},\mathbf{Z}_{1}^{\left(i\right)},w_{1}^{\left(i\right)}\right)\gets \left(\boldsymbol{\theta},\mathbf{Z},w\right)$
	
	\EndIf
	\EndWhile
	
	\item Resample $\left(\boldsymbol{\theta}^{1:N}_1,\mathbf{Z}^{1:N}_1\right)$ with weights $w^{1:N}_1$ and $w^{1:N}_1\gets \frac{1}{N}$
	
	\end{itemize}
	
	\For {$t:=2, \ldots, T$}
	
	$\left(\boldsymbol{\theta}_{t}^{1:N},\mathbf{Z}_{t}^{1:N}\right) \gets \left(\boldsymbol{\theta}_{t-1}^{1:N},\mathbf{Z}_{t-1}^{1:N}\right)$
	
	$w^{\left(i\right)}_{t}\gets \frac{\sum_{k=1}^{M}\mathbb{I}_{A_{t}}\left(\mathbf{z}_{t}^{\left(i,k\right)}\right)}{\sum_{k=1}^{M}\mathbb{I}_{A_{t-1}}\left(\mathbf{z}_{t}^{\left(i,k\right)}\right)}$, $i=1,\ldots,N$  
	
	resample $\left(\boldsymbol{\theta}^{1:N}_{t},\mathbf{Z}^{1:N}_{t}\right)$ with weights $w^{1:N}_{t}$ and $w^{1:N}_{t}\gets \frac{1}{N}$
	
	Sample $\boldsymbol{\theta}_{t}^{1:N}\sim K_{t}$
	\EndFor
	
	\Ensure  Particles $\boldsymbol{\theta}_{1:T}^{1:N}$.
	\end{algorithmic}
	\end{algorithm}

\subsubsection{Chaotic Stochastic Model}

Likelihood-based inference  breaks down for chaotic dynamics since small changes in the system parameters produce large changes in the system states later in time producing likelihoods which do not depend smoothly on the parameters \cite{Berliner1991}.  As an alternative, \cite{Wood10} propose an ABC related synthetic likelihood method based on a set of summary statistics that captures the important dynamics in the data rather than the noise-driven detail.  Using the Ricker map chaotic ecological\\ model  \citep{Turchin03, Wood10}, and some of our summary statistics from \cite{Wood10}, we employ the SCMC ABC algorithm, described above, to make inference about the model parameters.
	
	The scaled Ricker map describes the dynamics of a discrete population, $N_\nu$, over time as,
	\begin{equation*}
	N_{\nu+1}=rN_\nu\exp{\left(-N_\nu+e_\nu\right)},
	\end{equation*}
	where $e_\nu$ are independent normal errors with mean zero and variance, $\sigma^2_e$, that represent the process noise, and $r$ is the growth rate parameter. The data are the outcome of a Poisson distribution observed at $n=50$ time steps,
	\begin{equation*}
	\mathbf{y}\sim \mbox{Poisson}\left(\phi N_\nu\right),
	\end{equation*}
	where $\phi$ is a scaling parameter. The vector of parameters that inference is made about is given by $\boldsymbol{\theta}=\left(r,\sigma^2_e,\phi\right)^T$. The likelihood, $P\left(\mathbf{y}|\boldsymbol{\theta}\right)$, is obtained by integrating over $e_\nu$ and is therefore analytically and numerically intractable \citep{Davison03}, thereby, raising the demand for a likelihood-free approach. The summary statistics used in the SCMC ABC algorithm are, 
	\begin{align*}
	\boldsymbol{s}=\{&\mbox{med}(\mathbf{y}),\  \sum_{i=1}^n\frac{y_i}{n},\  \frac{\sum_{i=1}^n yI_{(1,\infty)}(y_i)}{\sum_{i=1}^nI_{(1,\infty)}(y_i) },\\& \sum_{i=1}^n I_{(10,\infty)}(y_i), \ \sum_{i=1}^n I_{\{0\}}(y_i),\\& \ Q_{0.75}(\mathbf{y}),\ \ \max(\mathbf{y})\}
	\end{align*}
	where $\text{med}\left(\mathbf{y}\right)$ is the median and $Q_{0.75}$ is the 75\% quantile. The distance measure used is 
\begin{equation*}
\rho\left(s\left(\mathbf{z}\right),s\left(\mathbf{y}\right)\right)=|s\left(\mathbf{z}\right)-s\left(\mathbf{y}\right)|.
\end{equation*} 
 The corresponding vector of tolerances used were chosen as 1\% quantiles of the distribution of the deviations $\rho(s(\mathbf{z}),s(\mathbf{y}))$ that were estimated by simulation; $\epsilon =(1, 1.88, 6.25, 1, 2, 10, 35)$. 
	
\begin{figure}[h!]
        \centering
        \begin{subfigure}[b]{0.22\textwidth}
                \centering
                \includegraphics[width=\textwidth]{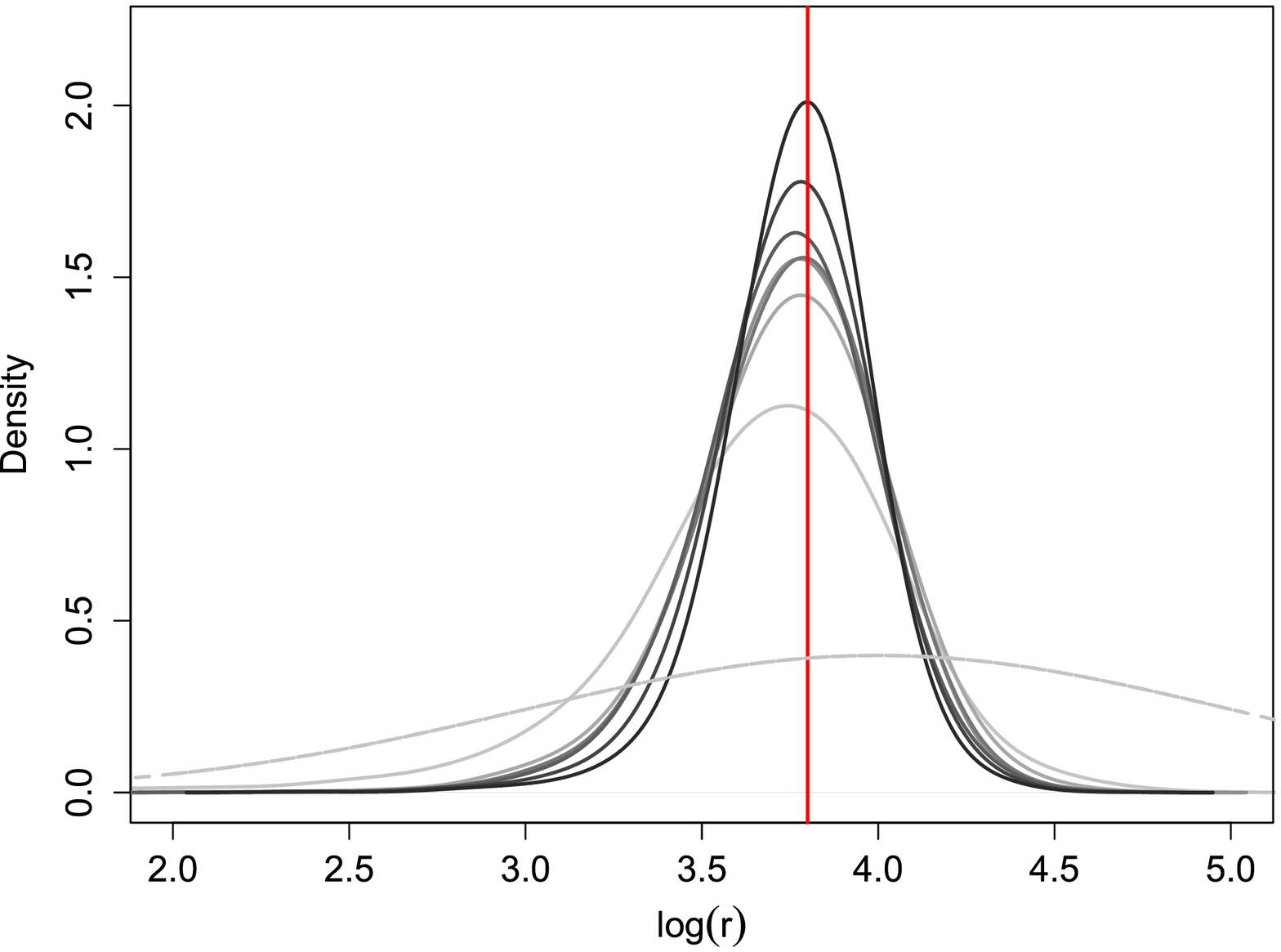}
                
                \label{prior_J}
        \end{subfigure}
        ~ 
        \begin{subfigure}[b]{0.22\textwidth}
                \centering
                \includegraphics[width=\textwidth]{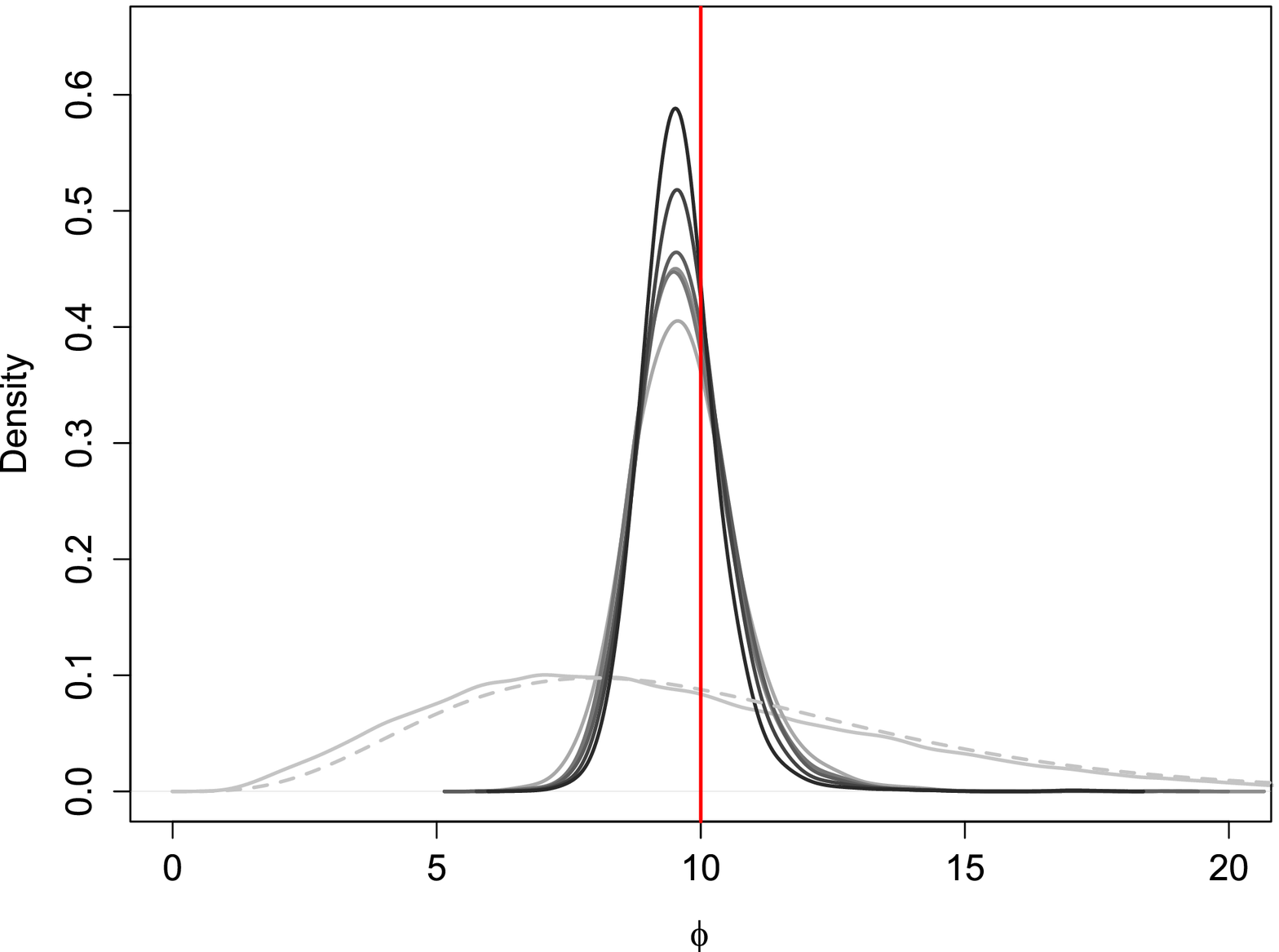}
                
                \label{post-J}
        \end{subfigure}
        ~
        \begin{subfigure}[b]{0.22\textwidth}
                \centering
                \includegraphics[width=\textwidth]{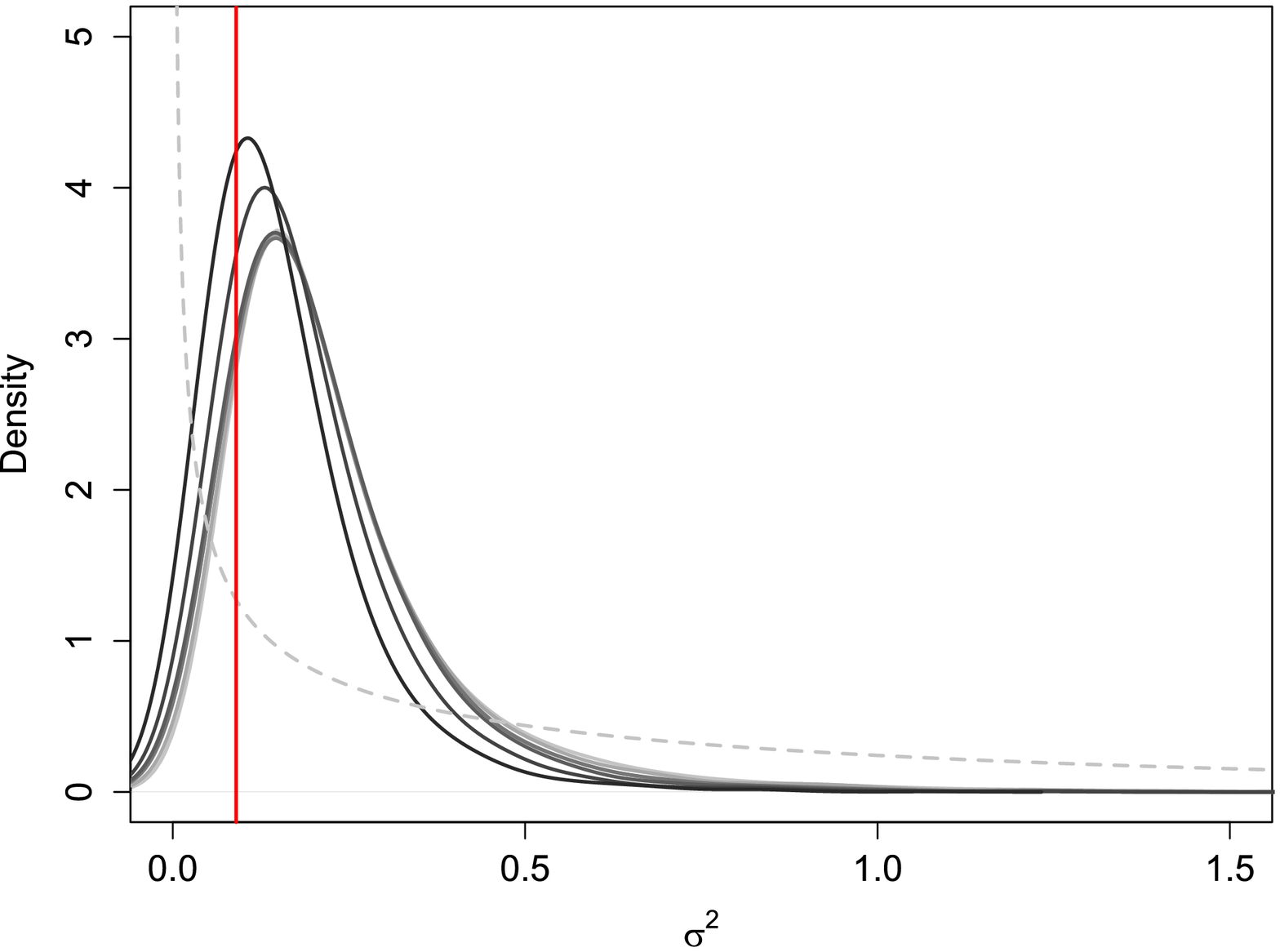}
               
                \label{eps-post}
        \end{subfigure}
        \caption{The Ricker model- kernel density estimates of the approximate marginal posteriors at times, $t=0,1,\ldots,T$, the color of the curves grows darker with time; the dashed, light gray curve is the prior density. The vertical lines are drawn at the true values of the parameters.}\label{post-kernel}
\end{figure}

\begin{figure}[h!]
        \centering
        \begin{subfigure}[b]{0.23\textwidth}
                \centering
                \includegraphics[width=\textwidth]{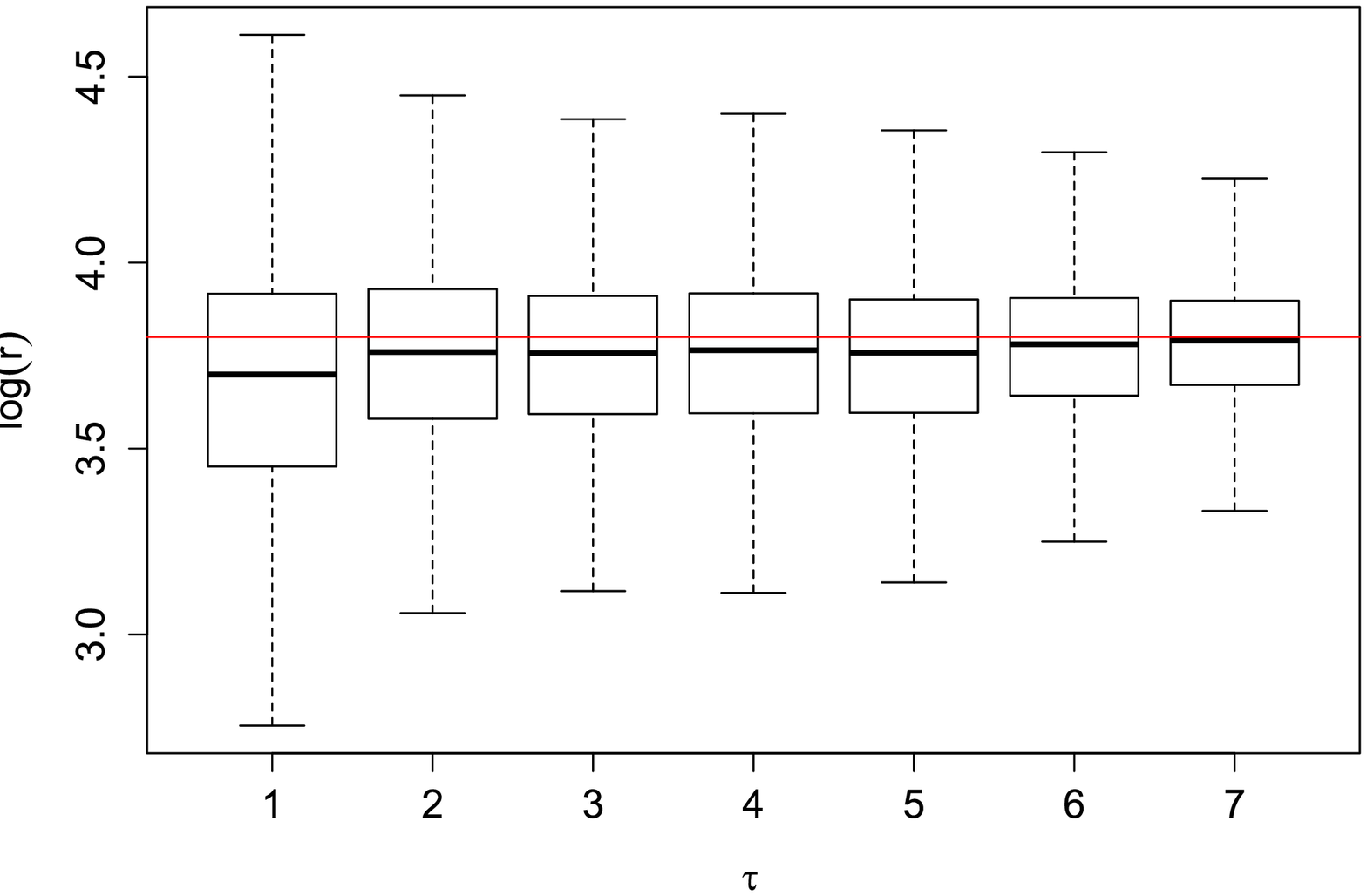}
                
                \label{prior_J}
        \end{subfigure}
        ~
        \begin{subfigure}[b]{0.23\textwidth}
                \centering
                \includegraphics[width=\textwidth]{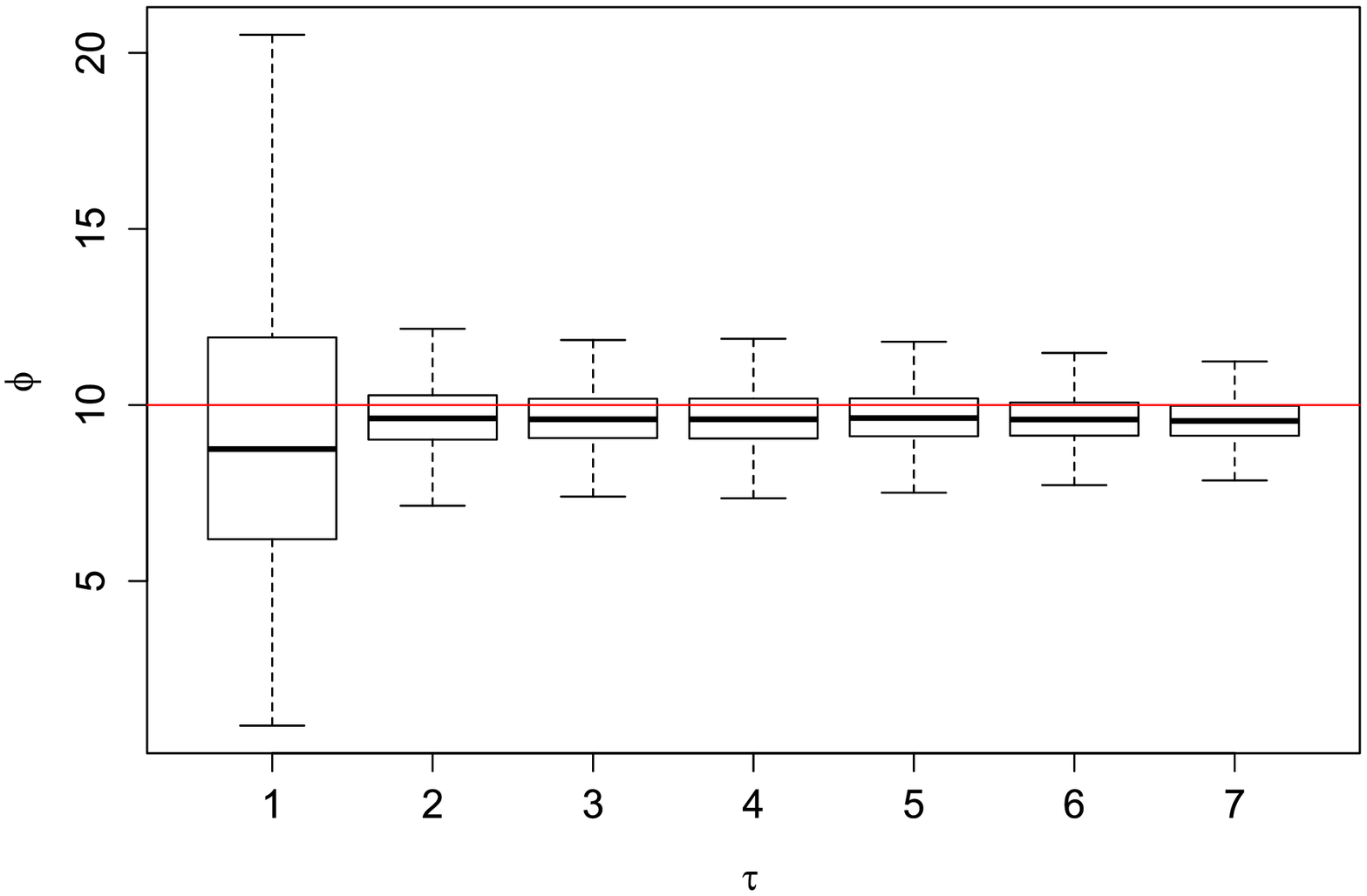}
                
                \label{post-J}
        \end{subfigure}
        ~
        \begin{subfigure}[b]{0.23\textwidth}
                \centering
                \includegraphics[width=\textwidth]{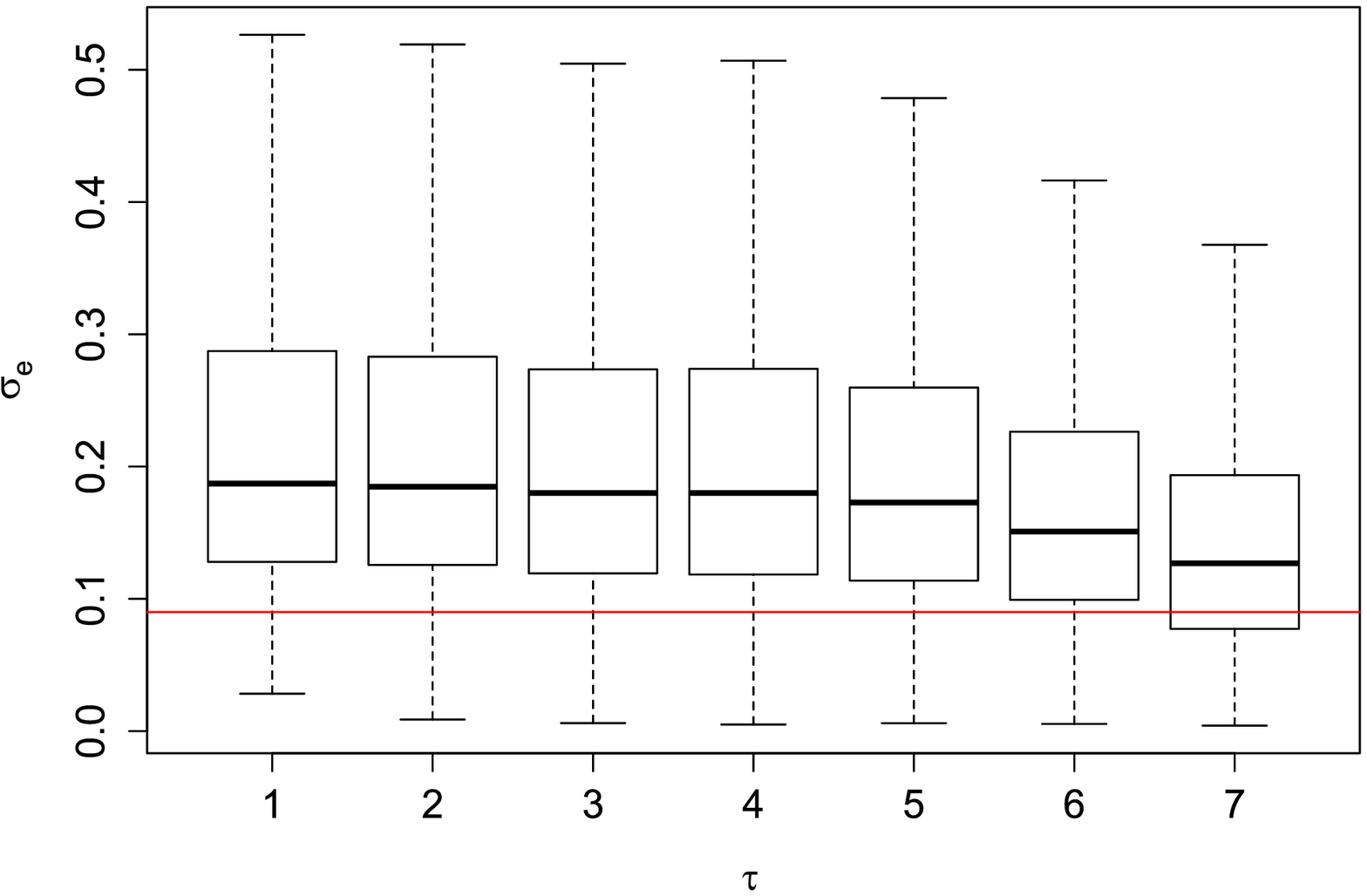}
               
                \label{eps-post}
        \end{subfigure}
     
        \caption{The Ricker model- approximate posterior box 
				plots evolving by sequential addition of summary statistics; the horizontal line is drawn at the true values of the parameters}\label{post-box}
\end{figure}

		Algorithm~\ref{scmc-abc} is used to sample from the joint posterior based on data simulated from\\ $\boldsymbol{\theta}=\left(\exp\left(3.8\right),10,0.09\right)$ as per \cite{Wood10}. The prior distributions are defined independently over the components of $\boldsymbol{\theta}$ as a log-Gaussian distribution  over $r$ with mean 4 and variance 1, a chi-squared distribution with 10 degrees of freedom for $\phi$ and an inverse gamma distributions with shape parameter 3 and scale parameter 0.5 for $\sigma^2_e$.   Sampling $\boldsymbol{\theta}_t^{1:N}$ from $K_t$ was performed through a Metropolis Hastings step using a $\chi^2_\theta$ transition density.	The number of summary statistics determines the number of steps taken in the SCMC ABC algorithm since we enter only one summary statistic at each time step, i.e., $T=7$. The results are presented in Figure~\ref{post-kernel} as kernel density estimates of the approximate marginal posteriors at the seven time steps and in Figure~\ref{post-box} as the marginal posterior box plots together with the true parameter values. The transition of the particles towards high probability regions and therefore focusing about the true value by addition of more summary statistics is evident from the plots.

With any ABC analysis, the quality of the parameter estimates depend primarily on the ability of $\boldsymbol{s}$ to summarize the important features of the data and the model.  When using Algorithm~\ref{scmc-abc}, the specific order of inclusion of those summaries into $A_t$ can make the algorithm easier or harder to apply due to its potential for particle degeneracy.  If addition of $s_t(\cdot)$ to $A_t$ induces a significant shift to the posterior, few or no particles remain with positive weights.  Figure~\ref{post-box} showcases the occasionally dramatic impact of summary inclusions and therefore the potential for large proportions of near zero resampling weights.

Setting the order of summary statistic inclusions can be done by maximizing the correlation between the summary statistics and/or pilot runs.  In cases of severe particle degeneracy, inclusion of summary statistics may need to be combined with a decreasing schedule for $\epsilon$ to control the effective sample size in practice.  However, in this example no manipulations of order or $\epsilon$ were necessary.

\section{Discussion}\label{sec:conclusion}
This paper introduces a sequential constraint based approach to defining the sequence of bridging distributions in SMC where part of the novelty of the algorithm is in facilitating posterior sampling through flexible definitions of constraints.  While some constraints have clear and interpretable impacts on the parameter space, such as parameter inequalities, model constraints often can not be readily transformed into constraints on the parameter space.  Consequently imposition of the constraint produces potentially serious challenges in posterior sampling.  In this paper we have proposed a new variant of the SMC samplers that can be used in the case that imposition of a constraint creates challenges in sampling from the target distribution. By defining the sequence of distributions using the specific parametrization of the constraints in each case we sequentially increase the rigidity of the constraint and through weighting, resampling and sampling steps obtain a sample from the fully constrained target distribution.   In Section~\ref{sec:probit_general} we proposed a general starting constraint enforcing strategy but there may be gains in redefining general applications as though they were built with constraints as was done in Section~\ref{sec:app_constraints} with the ODE and ABC models.  SCMC expands the SMC framework for such application specific sampling modifications.

Our examples illustrate the variety of frameworks in which the SCMC algorithm can be used. This wide scope of application is due to our broad interpretation of constraints; any restriction over the parameter space or the model that could be imposed through a number of intermediate steps can be assembled in the SCMC to provide the means of efficient posterior sampling.  Consequently the SCMC approach provides a non-unique path between the prior (or a computationally tractable distribution) and the posterior based on a non-unique set of constraints for any problem.  

In Section \ref{scmc-toy} the functional constraint is satisfied point-wise with probability 1 as $\tau\rightarrow\infty$ but is also satisfied with high probability with finite $\tau$.   However, in Section \ref{sec:manifold}, the constraint is satisfied with probability 0 with finite $\tau$ despite the degeneration of the density to the manifold.  In many cases this is sufficient to examine properties of the resulting posterior, but in other cases, full enforcement of the constraint is possible.   Consequently Sections \ref{sec:ode} and \ref{sec:abc} demonstrate alternative constraint inducing strategies. A sequence of probit functions designed to fully enforce the constraint in the limit as $\tau\rightarrow\infty$ is a general constraint inducing strategy, but sometimes alternative strategies are easier to implement.    

In Section \ref{scmc-toy} the path was defined so as to make use of the simplicity of sampling from the unconstrained regression model.  In Section \ref{sec:ode} a model relaxation constraint was introduced to simplify the topological difficulties associated with ODE models with a mixture of discrete and continuous parameters.  Alternative constraint strategies for the ODE model include relaxing the discrete nature of $I(0)$, or using an alternative smoothing based analogue to the smooth functional tempering approach of \cite{CampbellSteele2011}.  Our model relaxation technique was chosen due to the computational speed of the kernel smoothing step and obtaining the numerical solution to the ODE.  In Section \ref{sec:abc}, the SCMC was defined based on constraining parameters to meet an increasing set of summary statistic based criteria, but one could similarly define a strategy based on increasing the strictness of meeting all summaries at once through a decreasing set of $\epsilon$.  While both of these strategies hold advantages, one may prefer to choose a hybrid approach simultaneously adding new summaries and decreasing $\epsilon$ for previously included summaries.


%



\bibliographystyle{spbasic}      
\bibliography{refs}   

\begin{thebibliography}{31}
\providecommand{\natexlab}[1]{#1}
\providecommand{\url}[1]{{#1}}
\providecommand{\urlprefix}{URL }
\expandafter\ifx\csname urlstyle\endcsname\relax
  \providecommand{\doi}[1]{DOI~\discretionary{}{}{}#1}\else
  \providecommand{\doi}{DOI~\discretionary{}{}{}\begingroup
  \urlstyle{rm}\Url}\fi
\providecommand{\eprint}[2][]{\url{#2}}

\bibitem[{Beaumont et~al(2009)Beaumont, Cornuet, Marin, and
  Robert}]{Beaumont09}
Beaumont M, Cornuet JM, Marin JM, Robert C (2009) Adaptive approximate bayesian
  computation. Biometrika 96:983–990

\bibitem[{Berliner(1991)}]{Berliner1991}
Berliner LM (1991) Likelihood and bayesian prediction of chaotic systems.
  Journal of the American Statistical Association 86(416):938--952

\bibitem[{Beskos et~al(2014)Beskos, Crisan, and Jasra}]{Beskos14}
Beskos A, Crisan D, Jasra A (2014) On the stability of sequential monte carlo
  methods in high dimensions. Ann Appl Probab 24:1396--1445

\bibitem[{Box and Tiao(1973)}]{Box73}
Box GEP, Tiao GC (1973) Bayesian Inference in Statistical Analysis. Wiley

\bibitem[{Brunel(2008)}]{Brunel2008}
Brunel NJ (2008) Parameter estimation of odes via nonparametric estimators.
  Electronic Journal of Statistics

\bibitem[{Calderhead and Girolami(2009)}]{CalderheadGirolami}
Calderhead B, Girolami M (2009) Estimating bayes factors via thermodynamic
  integration and population mcmc. Computational Statistics and Data Analysis
  53:4028--4045

\bibitem[{Calderhead et~al(2009)Calderhead, Girolami, and
  Lawrence}]{CalderheadEtAl2009}
Calderhead B, Girolami M, Lawrence N (2009) Accelerating bayesian inference
  over nonlinear differential equations with gaussian processes. Advances in
  Neural Information Processing Systems

\bibitem[{Campbell and Steele(2011)}]{CampbellSteele2011}
Campbell D, Steele RJ (2011) Smooth functional tempering for nonlinear
  differential equation models. Statistics and Computing
  \doi{10.1007/s11222-011-9234-3}

\bibitem[{Campbell and Lele(2013)}]{Campbell13}
Campbell DA, Lele S (2013) An anova test for parameter estimability using data
  cloning with application to statistical inference for dynamic systems.
  Computational Statistics and Data Analysis
  http://dx.doi.org/10.1016/j.csda.2013.09.013

\bibitem[{Chopin(2002)}]{Chopin02}
Chopin N (2002) A sequential particle filter for static models. Biometrika
  89:539--552

\bibitem[{Davison(2003)}]{Davison03}
Davison AC (2003) Statistical Models. 94-160, 456-458, 605-619, Cambridge
  University Press

\bibitem[{Del~Moral et~al(2006)Del~Moral, Doucet, and Jasra}]{Moral06}
Del~Moral P, Doucet A, Jasra A (2006) Sequential monte carlo samplers. J R
  Statist Soc B 68:411–436

\bibitem[{Del~Moral et~al(2009)Del~Moral, Doucet, and Jasra}]{Moral09}
Del~Moral P, Doucet A, Jasra A (2009) An adaptive sequential monte carlo method
  for approximate bayesian computation. Tech. rep., University of British
  Columbia

\bibitem[{Dette et~al(2006)Dette, N., and Pilz}]{Dette06}
Dette H, N N, Pilz KF (2006) A simple nonparametric estimator of a strictly
  monotone regression function. Bernoulli 12:469--490

\bibitem[{Doucet et~al(2001)Doucet, De~Freitas, and Gordon}]{DoucetBook}
Doucet A, De~Freitas N, Gordon N (eds)  (2001) Sequential {M}onte {C}arlo
  Methods in Practice. Springer-Verlag

\bibitem[{Gelman et~al(1996)Gelman, Bois, and Jiang}]{GelmanBoisJiang1996}
Gelman A, Bois FY, Jiang J (1996) Physiological pharmacokinetic analysis using
  population modeling and informative prior distributions. Journal of the
  American Statistical Association 91(436):1400--1412

\bibitem[{He and Shi(1998)}]{He98}
He X, Shi P (1998) Monotone b-spline smoothing. Journal of the American
  Statistical Association 93:643--650

\bibitem[{Joyce and Marjoram(2008)}]{Joyce08}
Joyce P, Marjoram P (2008) Approximately sufficient statistics and bayesian
  computation. Statistical Applications in Genetics and Molecular Biology 7

\bibitem[{Liang and Wu(2008)}]{LiangWu2008}
Liang H, Wu H (2008) Parameter estimation for differential equation models
  using a framework of measurement error in regression models. Journal of the
  American Statistical Association 103

\bibitem[{Marin et~al(2012)Marin, Pudlo, Robert, and Ryder}]{Marin12}
Marin JM, Pudlo P, Robert CP, Ryder RJ (2012) Approximate bayesian
  computational methods. Statistics and Computing 22:1167--1180

\bibitem[{Massad et~al(2004)Massad, Coutinho, Burattini, and Lopez}]{Massad04}
Massad E, Coutinho F, Burattini M, Lopez L (2004) The eyam plague revisited:
  did the village isolation change transmission from fleas to pulmonary? Med
  Hypotheses 63:911–915

\bibitem[{Moffa and Kuipers(2014)}]{Moffa14}
Moffa G, Kuipers J (2014) {Sequential Monte Carlo EM for multivariate probit
  models.} Computational Statistics and Data Analysis 72:252–272

\bibitem[{Raggett(1982)}]{Raggett1982}
Raggett GF (1982) Modelling the eyam plague. Institue of Mathematics and its
  Applications 18:221--226

\bibitem[{Ramsay et~al(2007)Ramsay, Hooker, Campbell, and Cao}]{RamsayEtAl2007}
Ramsay J, Hooker G, Campbell DA, Cao J (2007) Parameter estimation for
  differential equations: a generalized smoothing approach. Journal of the
  Royal Statistical Society B 69

\bibitem[{Ramsay(1998)}]{Ramsay98}
Ramsay JO (1998) Estimating smooth monotone functions. Journal of the Royal
  Statistical Society Series B 60:365--375

\bibitem[{Ratmann et~al(2009)Ratmann, Andrieu, Wiuf, and
  Richardson}]{Ratmann09}
Ratmann O, Andrieu C, Wiuf C, Richardson S (2009) Model criticism based on
  likelihood-free inference, with an application to protein network evolution.
  Proceedings of the National Academy of Sciences of the United States of
  America 106:10,576--10,581

\bibitem[{Riihimaki and Vehtari(2010)}]{RV2010}
Riihimaki J, Vehtari A (2010) Gaussian processes with monotonicity information.
  Journal of Machine Learning Research: Proceedings of the Thirteenth
  International Conference on Artificial Intelligence and Statistics 9:645--652

\bibitem[{Sisson et~al(2009)Sisson, Fan, and Tanaka}]{Sisson09}
Sisson SA, Fan Y, Tanaka M (2009) Sequential monte carlo without likelihoods.
  Errata Proceedings of the National Academy of Sciences 106:16–889

\bibitem[{Tavare et~al(1997)Tavare, Balding, R.~C.~Griffiths, and
  Donnelly}]{Tavare97}
Tavare S, Balding DJ, R~C~Griffiths JRC, Donnelly P (1997) Inferring
  coalescence times from dna sequence data. Genetics 145:505--518

\bibitem[{Turchin(2003)}]{Turchin03}
Turchin P (2003) Complex Population Dynamics. 8-11, 47-77, Princeton University
  Press

\bibitem[{Wood(2010)}]{Wood10}
Wood SN (2010) Statistical inference for noisy nonlinear ecological dynamic
  systems. Nature 466:1102--1104

\end{thebibliography}

%
%

\end{document}